\documentclass[fleqn,twocolumn,10pt]{wlscirep}
\usepackage[utf8]{inputenc}
\usepackage[T1]{fontenc}

\usepackage{graphicx}
\usepackage{subfigure}
\usepackage{textcomp}
\usepackage{siunitx} 
\usepackage[version=4]{mhchem} 
\usepackage{hyperref}
\usepackage{amssymb}

\usepackage{lineno}

\usepackage{multibib}
\newcites{methods}{ }
\newcites{supp}{ }

\let\oldequation\equation
\let\oldendequation\endequation

\renewenvironment{equation}
  {\linenomathNonumbers\oldequation}
  {\oldendequation\endlinenomath}
 
\title{Impact of ionizing radiation on superconducting qubit coherence} 

\author[1,*]{Antti P.\ Veps\"al\"ainen}
\author[1]{Amir H.\ Karamlou}
\author[2,**]{John L.\ Orrell}
\author[1,4]{Akshunna S.\ Dogra}
\author[2]{Ben Loer}
\author[1]{Francisca Vasconcelos}
\author[3]{David K. Kim}
\author[3]{Alexander J. Melville}
\author[3]{Bethany M. Niedzielski}
\author[3]{Jonilyn L. Yoder}
\author[1]{Simon Gustavsson}
\author[1]{Joseph A.\ Formaggio}
\author[2]{Brent A.\ VanDevender}
\author[1,3]{William D.\ Oliver}

\affil[1]{Massachusetts Institute of Technology, Cambridge, MA 02139, USA}
\affil[2]{Pacific Northwest National Laboratory, Richland, WA 99352, USA}
\affil[3]{MIT Lincoln Laboratory, Lexington, MA 02421, USA}
\affil[4]{Harvard University, Cambridge, MA 02138, USA}

\affil[*]{Corresponding author for qubit operations: avepsala@mit.edu}
\affil[**]{Corresponding author for radiation exposure: john.orrell@pnnl.gov}

\begin{abstract}
\end{abstract}

\begin{document}

\flushbottom
\maketitle
\thispagestyle{empty}

%
%

{\bf 
The practical viability of technologies that rely on qubits requires long coherence times and high-fidelity operations~\cite{DiVincenzo2000}. Superconducting qubits are one of the leading platforms for achieving these objectives~\cite{Arute2019,Kandala2019}.
However, the coherence of superconducting qubits is impacted by broken Cooper pairs ~\cite{lutchyn2006,Martinis2009,Jin2015}, 
referred to as quasiparticles, whose experimentally observed density is orders of magnitude higher than the value predicted at equilibrium by the Bardeen-Cooper-Schrieffer (BCS) theory of superconductivity~\cite{Serniak:2018aa,Aumentado2004,Taupin2016}.
Previous work~\cite{Serniak2019,corcoles2011,barends2011} has shown that infrared photons significantly increase the quasiparticle density, yet even in the best isolated systems, it still remains much higher~\cite{Serniak2019} than expected, suggesting that another generation mechanism exists~\cite{Besbalov2016}. 
Here, we provide evidence that ionizing radiation from environmental radioactive materials and cosmic rays contributes to this observed difference. The effect of ionizing radiation leads to an elevated quasiparticle density, which we
predict would ultimately limit the coherence times of superconducting qubits of the type measured here to the millisecond regime. We further demonstrate that introducing radiation shielding reduces the flux of ionizing radiation and positively correlates with increased energy-relaxation time. Albeit a small effect for today’s qubits, reducing or otherwise mitigating the impact of ionizing radiation will be critical for realizing fault-tolerant superconducting quantum computers.
}

Over the past 20 years, superconducting qubit coherence times have increased more than five orders of magnitude due to improvements in device design, fabrication, and materials, from less than one nanosecond in 1999 \cite{nakamura1999} to more than \SI{100}{\micro\second} in contemporary devices~\cite{Oliver:2013aa,Kjaergaard2019}. 
Nonetheless, to realize the full promise of quantum computing, far longer coherence times will be needed to achieve the operational fidelities required for fault-tolerance~\cite{Gottesman1998}.
%

Today, the performance of superconducting qubits is limited in 
part by quasiparticles - a phenomenon known colloquially as ``quasiparticle poisoning.'' 
Although it was suggested~\cite{Grunhaupt:2018aa} and very recently confirmed~\cite{cardani2020reducing} that high-energy cosmic rays result in bursts of quasiparticles that reduce the quality factor in superconducting granular aluminum resonators, to date there has been no quantitative model or experimental validation of the effect of environmental ionizing radiation on superconducting qubits. 

In this work, we measure the impact of environmental radiation on superconducting qubit performance. 
We develop a model and determine its parameters by measuring the effect of radiation from a calibrated radioactive source on the qubit energy-relaxation rate. 
We use this model to infer the energy-relaxation rate $\Gamma_1 \approx \SI{1/4}{\per\milli\second}$ for our qubit 
if it were limited solely by the measured level of naturally occurring cosmic rays and background environmental radiation present in our laboratory. 
%
%
We then demonstrate that the deleterious effects of this external radiation can be reduced by protecting the device with a lead shield. 
The improvement in qubit energy-relaxation time from this independent shielding measurement is consistent with the radiation-limited $\Gamma_1$ inferred from the model.
Furthermore, we show that our estimate of the quasiparticle density due solely to the ionizing radiation agrees with the observed surplus quasiparticle density in qubits that are well-isolated from thermal photons~\cite{Serniak:2018aa,Serniak2019}.
This finding is of crucial importance for all superconducting applications 
in which quasiparticle excitations are harmful, such as superconducting quantum computing, superconducting detectors \cite{Day2003,Irwin1996}, or Majorana fermion physics~\cite{Albrecht2017}. 
%


\begin{figure*}[t!]
    \centering
    \includegraphics[width=\textwidth]{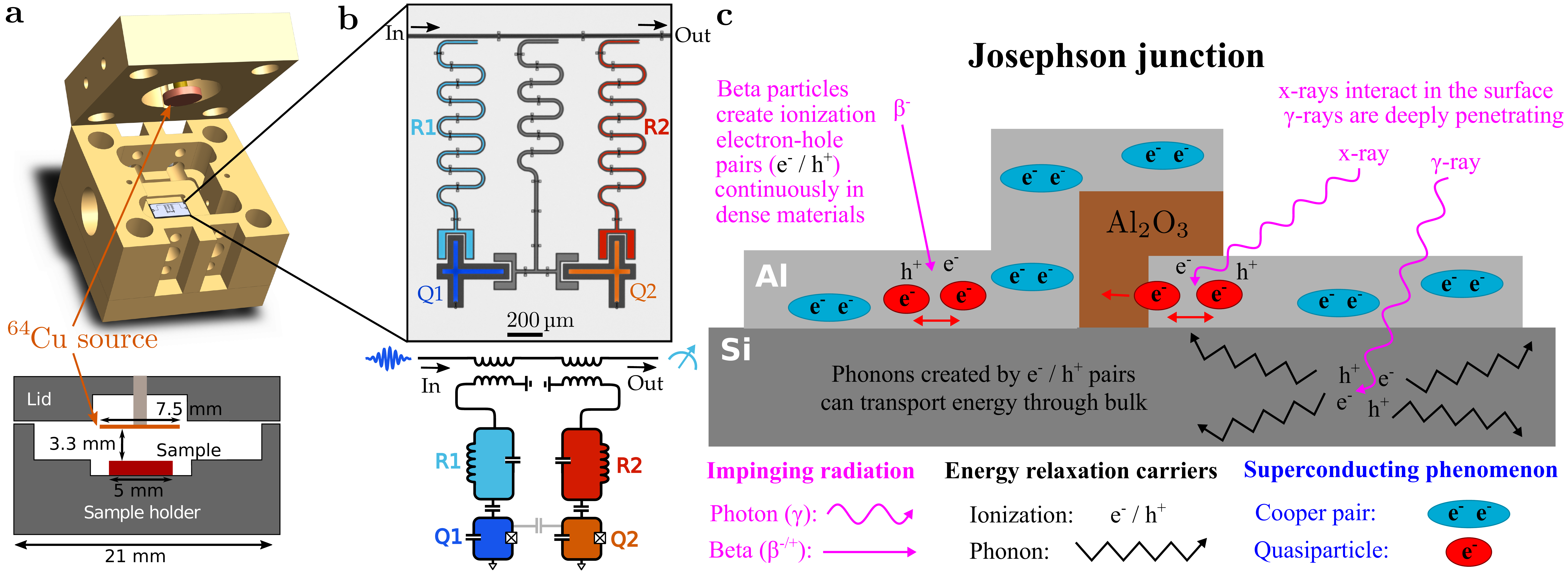}
    \caption{{\bf Schematic of the experiment.} 
    a) Illustration of the sample holder and the \ce{^64}{Cu} radiation source. The source is mounted \SI{3.3}{\milli\metre} above the silicon chip containing the superconducting aluminum transmon qubits. 
    b) False-color micrograph and circuit schematic of the qubit sample. The sample consists of two transmon qubits, Q1 (blue, left) and Q2 (orange, right). The resonators used to readout the qubits are shown with red and cyan. The resonators are inductively coupled to a common microwave transmission line, through which both qubit control and readout pulses are sent. The control pulses and the measurement pulses are generated using microwave sources and arbitrary waveform generators at room temperature (not shown, see Extended Data Fig. \ref{fig:ex1}a). 
    c) Diagram of the possible quasiparticle generation processes. Incoming ionizing radiation (from $\beta^{\pm}$, $\gamma$, and cosmic rays) interact with the Al qubit and Si substrate, creating electron-hole pairs due to the ionization of atoms and phonons (see text). The subsequent energy cascade of these particles ultimately breaks Cooper pairs and thereby generates quasiparticles. 
    }
    \label{fig:fig1}
\end{figure*}

For emerging quantum processors, one of most commonly used modalities 
is the superconducting transmon qubit~\cite{Koch2007a}, which 
comprises one or more Josephson junctions and a shunt capacitor. 
The intrinsic nonlinear inductance of the junction in combination with the linear capacitance
forms an anharmonic oscillator~\cite{Krantz2019}. 
The non-degenerate transition energies of such an 
oscillator 
are uniquely addressable, and in particular, 
its ground and first excited states 
serve as the logical $|0\rangle$ and $|1\rangle$ states of the qubit, respectively. 
In an ideal situation, 
qubits would suffer no loss of coherence during the 
the run-time of a quantum computation. 
However, interactions with the environment 
introduce decoherence channels, 
which for the case of energy decay, result in a loss of qubit polarization over time,
\begin{equation}
    \label{eq:coherence}
    p(t) = {\rm e}^{-\Gamma_1 t},
\end{equation}
where $p(t)$ is the excited-state probability and $\Gamma_1 \equiv 1/T_1$ is the energy relaxation rate corresponding to the relaxation time $T_1$, which limits the qubit coherence time. 
For such processes, the total energy relaxation rate is a combination of all individual rates affecting the qubit,
%
%
\begin{equation}
    \label{eq:gamma_1}
    \Gamma_1 = \Gamma_{\rm qp} + \Gamma_{\rm other},
\end{equation}
where $\Gamma_{\rm qp}$ is the energy relaxation rate due to the quasiparticles and $\Gamma_{\rm other}$ contains all other loss channels, such as radiation losses, dielectric losses, and the effect of two-level fluctuators in the materials~\cite{Klimov2018}. 
In the transmon, the quasiparticle energy-relaxation rate $\Gamma_{\rm qp}$  depends on the normalized quasiparticle density $x_{\rm qp} = n_{\rm qp} / n_{\rm cp}$ and the frequency of the qubit $\omega_q$, such that~\cite{Wang2014}
\begin{equation}
\label{eq:gamma_qp}
\Gamma_{\rm qp} = \sqrt{\frac{2\omega_q\Delta}{ \pi^{2}\hslash}} x_{\rm qp}.
\end{equation}
The Cooper pair density $(n_{\rm cp})$ and the superconducting gap $(\Delta)$ are material-dependent parameters, and for thin-film aluminum they are $n_{\rm cp} \approx \SI{4e6}{\per\cubic\micro\metre}$ and $\Delta \approx\SI{180}{\micro\electronvolt}$. 
This relation allows us to use the { energy-relaxation time} of a transmon as a sensor for quasiparticle density in the superconductor as well as to estimate the maximum { energy-relaxation time} of a transmon given a certain quasiparticle density. 
The thermal equilibrium contribution to $x_{\rm qp}$ is vanishingly small at the effective temperature of the sample, $T_{\rm eff} \approx \SI{40}{\milli\kelvin}$, compared with the other 
generation mechanisms we shall consider here. 

Currently, there exists no quantitative microscopic model directly connecting interactions of ionizing radiation (\textit{e.g.,} betas, gammas, x-rays, etc.) to quasiparticle populations in superconductors. 
However, a phenomonological picture describing the processes involved in this connection is shown in Fig.~\ref{fig:fig1}c.
The energy of ionizing radiation absorbed in the aluminum metal and silicon substrate is initially converted into 
ionization electron-hole pairs.
%
We purposefully distinguish these high-energy excitations due to the ionization of atoms -- which occur in both aluminum and silicon -- from the lower-energy quasiparticle excitations resulting from broken Cooper-pairs in aluminum.  
Thereafter, a non-equilibirum relaxation cascade involving secondary ionization carrier and phonon production serves to
transfer the absorbed radiation power to and within the aluminum qubit, where it breaks Cooper pairs and generates quasiparticles \cite{Kozorezov2000, kozorezov2007electron}. 
%

To estimate the effect of the radiation intensity measured in the laboratory, we employ a radiation transport simulation (see Methods for details) to calculate the total quasiparticle-generating power density $P_{\rm tot}$ close to the qubit due to radiation sources. We use a simple model for quasiparticle dynamics,~\cite{Wang2014}
%
\begin{equation}\label{eq:simple_qp_model}
    \dot{x}_{\mathrm{qp}}(t) = -r x_{\mathrm{qp}}^{2}(t) - s x_{\mathrm{qp}}(t) + g,
\end{equation}
where $g$ is the quasiparticle generation rate, which linearly depends on $P_{\rm tot}$, $r$ is the recombination rate, and $s$ is the quasiparticle trapping rate. A steady state solution for the quasiparticle density is given by $x_{\mathrm{qp}} = (-s+\sqrt{s^2+4rg})/2r$, and if quasiparticle trapping is neglected ($s=0$), then $x_{\mathrm{qp}} = \sqrt{g/r}$. In a separate quasiparticle injection experiment, we verified that this is a valid approximation in our devices, see Extended Data Fig. \ref{fig:ex3} and Supplementary material for discussion. By substituting the model for $x_{\mathrm{qp}}$ into 
Eq.~\eqref{eq:gamma_qp} and using Eq.~\eqref{eq:gamma_1}, the qubit decay rate is given by
\begin{equation}
\label{eq:gamma_power}
    \Gamma_1 = a\sqrt{\omega_{\rm q} P_{\rm tot}} + \Gamma_{\rm other},
\end{equation}
where $a$ is a coefficient accounting for unknown material parameters and the conversion from absorbed radiation power to quasiparticle generation rate. In addition to the materials of the chip, the conversion efficiency depends on the phononic losses and the thermalization of the sample. The value of $a$ can be experimentally determined by exposing the qubit to a known source of ionizing radiation.



\section*{Results}

\begin{figure}[!t]
    \centering
    \includegraphics[width=\columnwidth]{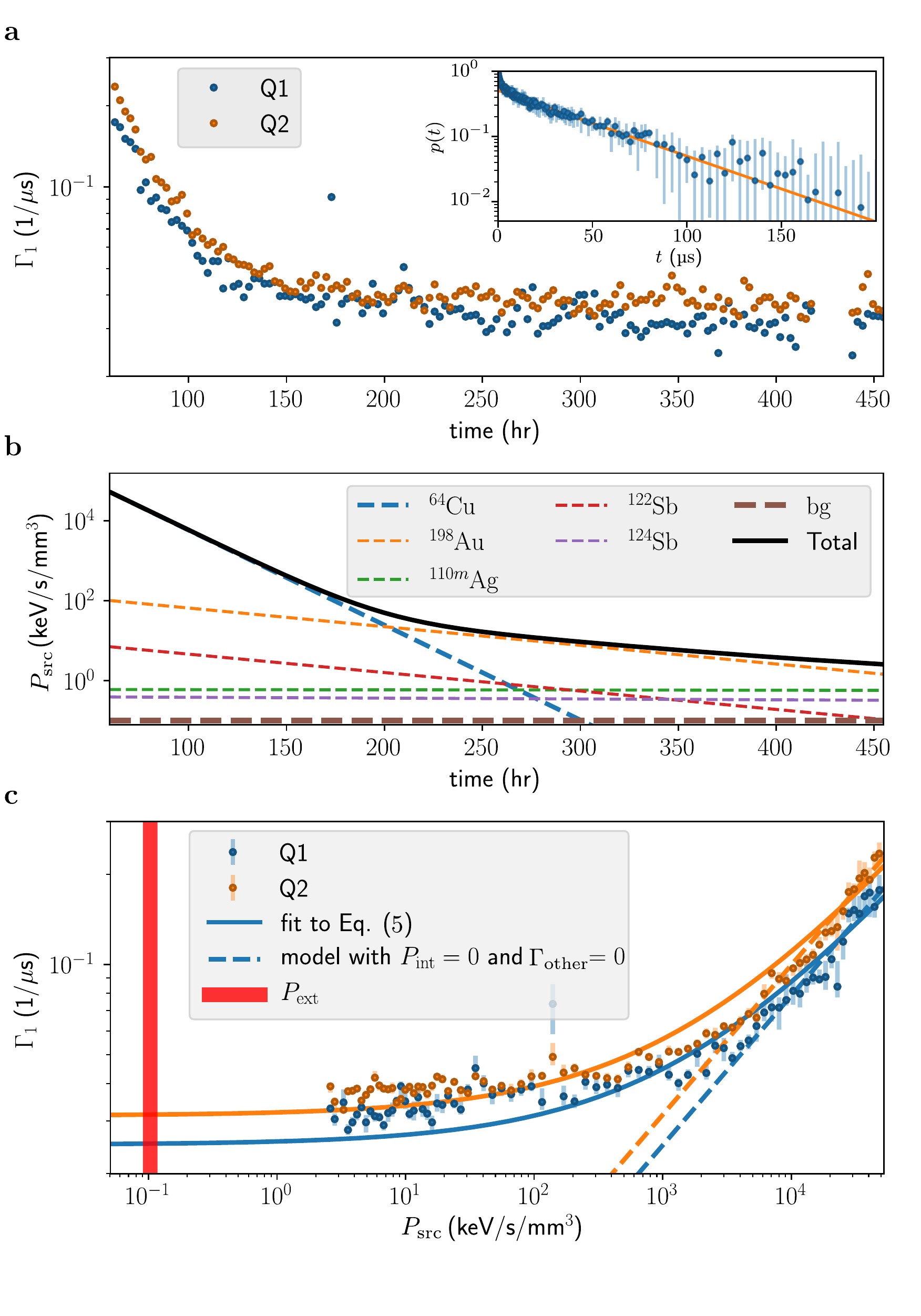}
    \caption{{\bf \ce{^64}{Cu} radiation exposure experiment.} 
    a) Measured energy relaxation rates $\Gamma_1 = 1/T_1$ of qubits Q1 (blue) and Q2 (orange) as a function of time when exposed to the \ce{^64}{Cu} source. The inset shows an example of the raw data used for fitting the energy relaxation rates. Blue points are the median of 20 measured qubit excited-state populations $p(t)$ at various times after the excitation pulse. Blue bars indicate the 95\% confidence interval for the median. The orange line is the exponential fit to the data, given in Eq.~\eqref{eq:coherence}. 
    The super-exponential decay at short measurement times results from 
    statistical fluctuations in the quasiparticle-induced
    energy-relaxation rate during the 20 measurements~\cite{Gustavsson1573}.
    b) Power density of the radiation during the experiment derived from radiation transport simulations (see text). 
    c) Energy relaxation rates $\Gamma_1$ as a function of radiation power density. The solid lines show the fit to the model of Eq.~\eqref{eq:gamma_power}. The dashed lines show the fit to model of Eq.~\eqref{eq:gamma_power} with $\Gamma_{\rm other} = 0$ and $P_{\rm int} = 0$. The vertical red line is the radiation power density level due to the external radiation $P_{\rm ext}$.
    }
    \label{fig:fig2}
\end{figure}
\subsection*{Radiation exposure experiment}
To quantify the effect of ionizing radiation on superconducting qubits and to measure the coefficient $a$ in Eq.~\eqref{eq:gamma_power}, we inserted a \ce{^64}{Cu} radiation source 
%
in close proximity to a chip
containing two transmon qubits, Q1 and Q2, with average energy-relaxation rates of  $\Gamma_1^{(\rm Q1)} = \SI{1/40}{\per\micro\second}$ and $\Gamma_1^{(\rm Q2)} = \SI{1/32}{\per\micro\second}$, and transition frequencies $\omega_{\rm q}^{(\rm Q1)} = \SI[product-units=single]{2\pi x 3.48}{\giga\hertz}$ and $\omega_{\rm q}^{(\rm Q2)} = \SI[product-units=single]{2\pi x 4.6}{\giga\hertz}$, see Figs.~\ref{fig:fig1}a and~\ref{fig:fig1}c. \ce{^64}{Cu} has a short half-life of \SI{12.7}{\hour}, which permits an observation of the transition from elevated ionizing radiation exposure to normal operation conditions within a single cooldown of the dilution refrigerator. \ce{^64}{Cu} was produced by irradiating high-purity copper foil in the MIT Nuclear Reactor Laboratory (see Methods for details).

The energy relaxation rate $\Gamma_1$ of both qubits was repeatedly measured for over 400 hours during the radioactive decay of the \ce{^64}{Cu} source (see Fig.~\ref{fig:fig2}a, Methods and Extended Data Fig. \ref{fig:ex7}b). 
During this interval of time, the energy relaxation rate $\Gamma_1^{(\rm Q1)}$ of Q1 
decreased from \SI{1/5.7}{\per\micro\second}  to \SI{1/35}{\per\micro\second} due to the gradually decreasing radioactivity of the source, and similarly for Q2. 
The half-life was long enough to measure individual $\Gamma_1$ values at essentially constant levels of radioactivity, yet short enough to sample $\Gamma_1$ over a wide range of radiation powers, down to almost the external background level. 
In addition to affecting qubit energy-relaxation time, the resonance frequencies $\omega_{\rm r}$ of the readout resonators shifted due to quasiparticle-induced changes in their kinetic inductance, consistent with the quasiparticle recombination model of Eq.~\eqref{eq:simple_qp_model}. Similarly, we observed a slight shift in the qubit frequencies and a reduced $T_2$ time (see Supplementary material and Extended Data Figs. \ref{fig:ex4} and \ref{fig:ex5}).

The intensity of the radiation source used in the experiment was calibrated as a function of time using the gamma-ray spectroscopy of a reference copper foil that had been irradiated concurrently. 
%
The foils included a small amount of longer-lived radioactive impurities that began to noticeably alter the radiated power density expected for \ce{^64}{Cu} 
about 180 hours into the measurements (see Fig.~\ref{fig:fig2}b). 
For both the \ce{^64}{Cu} and the long-lived impurities, the radiation intensities from the different isotopes were converted to a single ionizing radiation power density using the radiation transport simulation package Geant4~\cite{1610988,AGOSTINELLI2003250} (see Methods for details). 
The contributions of the different isotopes (dashed lines) and the resulting net power density (solid line) of the radiation from the source, $P_{\rm src}$, are shown in Fig.~\ref{fig:fig2}b over the measurement time window. 
%
%

Using the data in Fig. \ref{fig:fig2}b as a method for calibrated time-power conversion, the energy relaxation rates of qubits Q1 and Q2 
are presented as a function of the radiation power density $P_{\rm src}$ (Fig.~\ref{fig:fig2}c). 
In the high-$P_{\rm src}$ limit (short times), the model of Eq.~\eqref{eq:gamma_power} can be fit to the data to extract the value for the conversion coefficient $a = \SI{5.4e-3}{\sqrt{\milli\metre^3\per\kilo\electronvolt}}$ by assuming $P_{\rm tot} \approx P_{\rm src}$ dominates all radiation sources that generate quasiparticles as well as all other decay channels. 
In the low-$P_{\rm src}$ limit (long times), the qubit { energy-relaxation rate} is limited predominantly by the decay rate $\Gamma_{\rm other}$ and, to a lesser extent, by the long-lived radioactive impurities in the foil.

%
%
%
Having determined the coefficient $a$ in Eq.~\eqref{eq:gamma_power}, we now remove the calibrated radiation source.
In its absence, the total radiation power density that generates quasiparticles can be categorized into two terms, $P_{\rm tot} = P_{\rm int} + P_{\rm ext}$.
The term $P_{\rm int}$ 
accounts for radiation power sources that are internal to the dilution refrigerator, such as infrared photons from higher temperature stages 
or radioactive impurities present during the manufacturing of the refrigerator components. 
$P_{\rm ext}$ is the external ionizing radiation source outside the dilution refrigerator whose influence on the qubits we attempt to determine in the shielded experiment described in the next section. 

\begin{figure}[t!]
\includegraphics[width=\columnwidth]{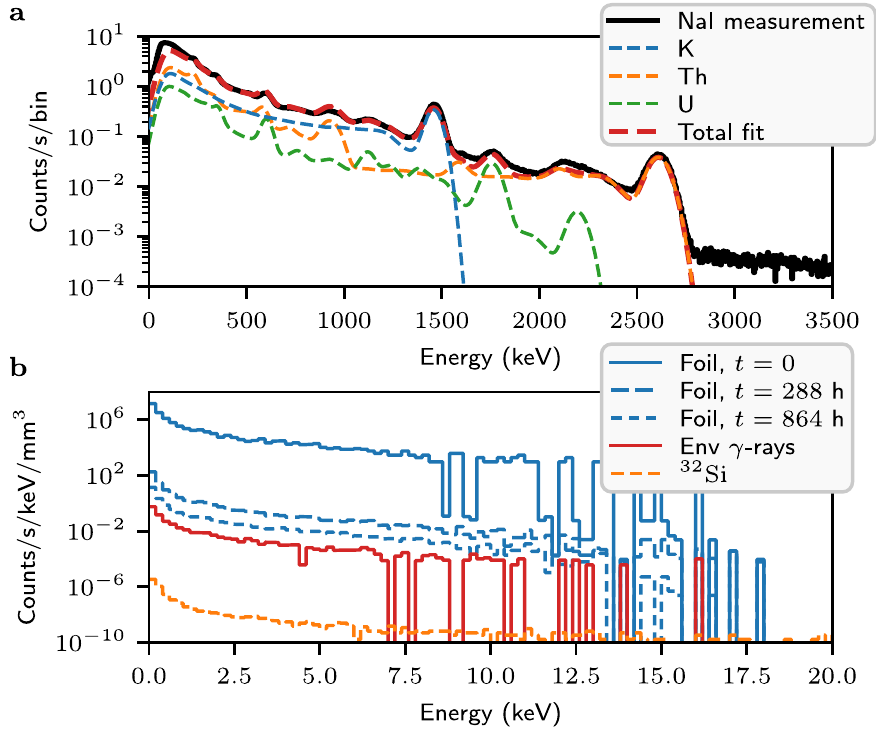}
\caption{{\bf Environmental radioactivity assessment.} {a) Spectrum of $\gamma$-radiation in the laboratory measured with NaI scintillation detector, binned in 8192 energy bins. The data is fit to the weighted sum of simulated spectra from \ce{^238}{U},\ce{^232}{Th}, and \ce{^40}{K} progenitors convolved with a response function of the NaI detector. These isotopes are typical contaminants in concrete.
b) Simulated spectral density of power absorbed in the aluminum film that comprises the qubit, calculated with Geant4 using the measured spectrum shown in panel a) and the emission spectra of the \ce{^64}{Cu} source and its impurities. At $t=0$, the spectrum is dominated by \ce{^64}{Cu}, after 12 days by \ce{^198}{Au} impurities, and after 36 days by \ce{^{\rm 110m}}{Ag}. \ce{^32}{Si} is a radioactive contaminant intrinsic to the silicon substrate~\cite{Aguilar-Arevalo:2016ndq}. The fluctuations in the simulated spectra are due to finite simulation statistics.}
}

\label{fig:fig3}
\end{figure}

%
To estimate the contribution of external radiation power $P_{\rm ext}$ to the data shown in Fig.~\ref{fig:fig2},
%
we directly measured the energy from the radiation fields present in the laboratory arising from $\gamma$-rays (see Fig. \ref{fig:fig3}) and cosmic rays, including those due to secondary processes, such as muon fields, using a NaI radiation detector (see Methods and Extended Data Fig. \ref{fig:ex2}e).
The spectra were used to determine the radiation intensities from cosmic rays and naturally occurring radioactive isotopes in the laboratory. 
These measured intensities were then used in a Geant4 radiation transport simulation to estimate the total external power density $P_{\rm ext} = (0.10 \pm 0.02) \si{\kilo\electronvolt\per\cubic\milli\meter\per\second}$
deposited in the aluminum 
that constitutes the resonators and qubits.
About 60\% of the external radiation power density results from the radioactive decays within the concrete walls of the laboratory (\SI{0.06}{\kilo\electronvolt\per\cubic\milli\metre\per\second}), with cosmic rays contributing the remaining 40\% (\SI{0.04}{\kilo\electronvolt\per\cubic\milli\metre\per\second}). 
This external power level is indicated with a vertical red band in Fig. \ref{fig:fig2}c. 
Although statistical errors in the measured intensities are small, we find a combined systematic uncertainty of approximately 20\%.  
The different sources' contributions to the total systematic uncertainty are detailed in the Methods section. 

Using the model in Eq.~\eqref{eq:gamma_power} with the determined parameters for $a$ and $P_{\rm ext}$ and the known qubit frequencies, the lower limit on the total energy relaxation rate due to the external radiation $P_{\rm ext}$ in the absence of all other energy-relaxation mechanisms is $\Gamma_1^{\rm (Q1)} \approx \SI{1/3950}{\per\micro\second}$ and $\Gamma_1^{\rm (Q2)} \approx \SI{1/3130}{\per\micro\second}$, 
corresponding to where the dashed lines would intersect the vertical red band in Fig.~\ref{fig:fig2}c.
These rates correspond to the point at which naturally occurring radiation from the laboratory would become the dominant limiting contributor to the qubit { energy-relaxation rate}. 
Although its effect on the energy-relaxation time is not dominant for today's qubits, ionizing radiation will need to be considered when aiming for coherence times required for fault-tolerant quantum computing. 
We can furthermore apply Eq.~\eqref{eq:gamma_qp} to estimate the quasiparticle density caused by the ionizing radiation background, giving $x_{\rm qp} \approx \num{7e-9}$, which agrees well with the lowest reported excess quasiparticle densities \cite{Serniak2019}.

\subsection*{Shielding experiment}
\begin{figure}[!t]
    \centering
    \includegraphics[width=0.95\columnwidth]{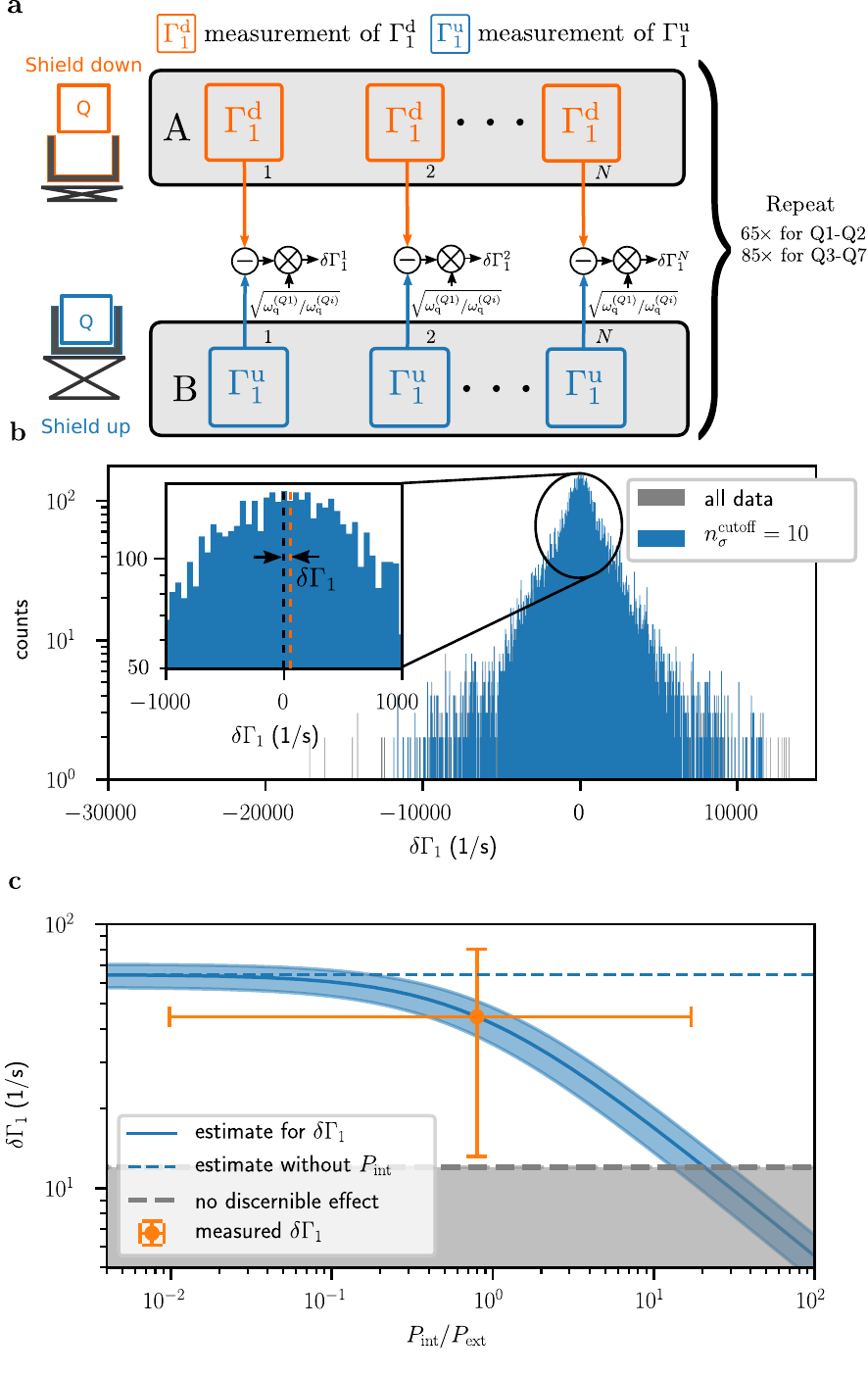}
    \caption{{\bf Qubit shielding experiment.} 
    a) Schematic of shielding experiment. The qubit energy relaxation rate is measured $N$ times with the shield up, and then again $N$ times with the shield down. This cycle is repeated 65 times for qubits Q1 and Q2 and 85 times for qubits Q3-Q7.
    %
    b) Histogram of the differences in energy relaxation rates when the shield is up versus down. The inset shows the histogram peak. The orange vertical line indicates the median of the distribution. Although the median difference in the relaxation rates between shield-up and shield-down configurations is only 1.8\% of the width of the distribution, it differs from zero in a statistically significant manner. 
    c) Difference in the energy relaxation rates in the shielding experiment (orange dot) versus $P_{\rm int} / P_{\rm ext}$. Vertical error bars are the 95\% confidence intervals for the median of $\delta\Gamma_1$. Horizontal error bars are the corresponding confidence intervals for $P_{\rm int}$. The blue line indicates the energy relaxation rate estimated using the model from the \ce{^64}{Cu} radiation exposure measurement and Eq. \eqref{eq:delta_gamma}. The filled blue region shows the confidence interval for the estimate assuming $\pm$ 20\% relative error for $P_{\rm ext}$. Below the gray dashed line, the experiment is not sensitive enough to detect $\delta\Gamma_1$.
    }
    \label{fig:fig4}
\end{figure}

We sought to verify the above result by shielding the qubits with \SI{10}{\centi\metre} thick lead bricks outside the cryostat to reduce the external radiation and thereby improve the qubit energy-relaxation times, see Fig.~\ref{fig:fig4}. 
The shield was built on a scissor lift to be able to cyclically raise and lower it to perform an A/B test of its effect.
%
%
By using the parameters extracted from the radiation exposure measurement and the model in Eq.~\eqref{eq:gamma_power}, the expected improvement of the energy relaxation rate due to the shield can be estimated from
\begin{align}
\label{eq:delta_gamma}
    &\delta \Gamma_1  
    \equiv \Gamma_1^{{\rm d}} - \Gamma_1^{{\rm u}} \notag\\
    &=a\sqrt{\omega_{\rm q}}\left(\sqrt{P_{\rm int} + (1 - \eta^{\rm d}) P_{\rm ext}} -\sqrt{P_{\rm int} + (1 - \eta^{\rm u}) P_{\rm ext}}\right),
\end{align}


\noindent where $\eta$ is the fraction of ionizing radiation blocked by the shield and the label u (d) corresponds to the parameters when the shield is up (down). %
%
%
We can make a realistic estimate of the efficiency of the shield by measuring the radiation energy spectrum with and without the shield using a NaI detector, giving $\eta^{\rm u} = 46.1 \%$. The shield blocks approximately 80\% of the radiation from the nuclear decay events in the laboratory, but is inefficient against the cosmic rays, see Methods for details.
From Eq.~\eqref{eq:delta_gamma}, in the absence of internal radiation sources 
($P_{\rm int} = 0$), the expected effect of the shield on the energy-relaxation rate of Q1 is
$\delta\Gamma_1 \approx \SI{1/15.5}{\per\milli\second}$, 
which is only 0.26 \% of the energy-relaxation rate of qubit Q1. 
%

To detect a signal this small, we measured the energy relaxation rates of the qubits while periodically placing the shield in the up and down positions and then comparing their difference over many cycles, similar in spirit to a Dicke switch radiometer measurement~\cite{Dicke1946}, see Fig~\ref{fig:fig4}a for a schematic. A single up/down cycle of the lead shield lasted 15 minutes.
To 
accelerate the data acquisition, we installed an additional sample in the dilution refrigerator with 5 qubits similar to Q1 and Q2. 

Fig.~\ref{fig:fig4}b shows the histogram of the accumulated differences in the energy relaxation rates, $\delta\Gamma_1$, for all of the qubits over the entire measurement, normalized to the frequency of Q1 using Eq.~\eqref{eq:gamma_power}.
From the median of the histogram, we estimate the shift in the energy relaxation rate $\delta\Gamma_1 = \SI{1/22.7}{\per\milli\second}$, 95\% CI [1/75.8,  1/12.4] \si{\per\milli\second}. 
The Wilcoxon signed-rank test can be used to assess if the measured energy-relaxation rates are lower in the shield up than in the shield down positions, and it yields a p-value of 0.006. 
As the p-value is much less than 0.05, we can reject the null hypothesis that the shield did not have any effect on the qubit energy-relaxation time with high confidence, see Methods and Extended Data Fig. \ref{fig:ex6} for additional details on the statistical analysis.

In Fig. \ref{fig:fig4}c, we have compared the result of the shielding experiment to the estimate of the effect of the background radiation obtained from the radiation exposure measurement. 
The orange dot shows $\delta\Gamma_1$ extracted from the shielding experiment. %
The solid blue line shows how this value would trend based on the predicted effect of the shield 
at the given $P_{\rm ext}$ measured in the laboratory for different values of internal radiation power density $P_{\rm int}$. 
Although we do not know the exact value of $P_{\rm int}$, we can approximate it by substituting the measured $\delta\Gamma_1$ and $a$ into Eq. \eqref{eq:delta_gamma} and by solving for $P_{\rm int} \approx \SI{0.081}{\kilo\electronvolt\per\cubic\milli\metre\per\second}$, 95\% CI [0, 1.73] \si{\kilo\electronvolt\per\cubic\milli\metre\per\second}. 
This value for $P_{\rm int}$, along with $P_{\rm ext}$, corresponds to a total quasiparticle density $x_{\rm qp} \approx \num{1.0e-8}$, again, consistent with earlier observations~\cite{Serniak2019}. 


Despite the 
uncertainty in the 
specific value of $P_{\rm int}$, the results acquired from the two independent experiments agree remarkably well. 
We conclude that, in the absence of all other energy-relaxation mechanisms, the ionizing radiation limits the qubit { energy-relaxation rate} to $\Gamma_1 \approx \SI{1/4}{\per\milli\second}$. 
In turn, shielding the qubits from environmental ionizing radiation improved their { energy-relaxation time}. 
The observed energy relaxation rate was reduced by
$\delta\Gamma_1 \approx \SI{1/22.7}{\per\milli\second}$, which is an 18 \% improvement of the radiation induced $\Gamma_1$ of the qubits. The shield was not able to remove all of the effects of the radiation, due to both the presence of internal radiation $P_{\rm int}$ and the imperfect efficiency of the shield.






\section*{Discussion}

The first reported results of the systematic operation of superconducting transmon qubits under intentionally elevated levels of ionizing radiation clearly show a deleterious effect on the performance of the qubits. We quantitatively determined the impact of radiation power density on the qubit  energy-relaxation time and showed that naturally occurring ionizing radiation in the laboratory creates excess quasiparticles in superconductors, reducing the qubit energy-relaxation time.

By employing shielding techniques commonly applied in neutrino physics and the search for dark matter,~\cite{Aguilar-Arevalo:2016ndq,PhysRevD.95.082002,Alduino:2018aa,Agostini:2018aa,Gando:2016aa,Aalseth:2018aa,Albert:2018ab} 
we improved the energy-relaxation rate of our qubits by approximately 0.2\%.
Although a rather small improvement for today's qubits, which are currently limited by other relaxation mechanisms, a simple model of the ionization generation of quasiparticles indicates that transmon qubits of this design will need to be shielded against ionizing radiation -- or otherwise redesigned to mitigate the impact of its resulting quasiparticles -- in order to reach energy relaxation times in the millisecond regime. Additionally, as was recently done with resonators \cite{cardani2020reducing}, locating qubit systems deep underground where cosmic rays and cosmogenic activation are drastically reduced should provide benefits for advancing quantum computing research.

Our results also shed light on a decades-old puzzle, namely, the origin of non-equilibrium quasiparticles observed broadly in experiments with superconductors at milliKelvin temperatures. Our measurements indicate that ionizing radiation accounts for a significant fraction of the residual excess quasiparticle levels observed in otherwise carefully filtered experiments, 
with impact on many fields that employ superconducting circuitry. For example, excess quasiparticles reduce the sensitivity of kinetic inductance detectors and transition edge sensors used in astronomy. Additionally, quasiparticle poisoning is a major impediment facing topologically protected Majorana fermions. Identifying ionizing radiation as a dominant source of excess quasiparticles is a first step towards developing techniques -- such as lead shielding, quasiparticle trapping, and designing devices with reduced quasiparticle sensitivity -- to mitigate its impact on superconducting circuits, including those used for quantum computation.

%
%
%
%

%

%
%



%
%

%
%

\bibliography{references}

%
%

%
%

\section*{Acknowledgements}


The authors thank Kyle Serniak and Roni Winik for many useful discussions and comments on the manuscript; Greg Calusine, Kyle Serniak, and Uwe von Luepke for designing and pre-characterizing the qubit samples; Grecia Castelazo for assistance with operating the lead shield; Mitchell S. Galanek, Ryan Samz, and Ami Greene for assistance in oversight of radiation source use; Michael R. Ames and Thomas I. Bork at the MIT Reactor (MITR) for production of the $^{64}$Cu source; Mital A. Zalavadia for providing the NaI detector.


This work was supported in part by the U.S. Department of Energy Office of Nuclear Physics under an initiative in Quantum Information Science research (Contract award \# DE-SC0019295, DUNS: 001425594); by the U.S. Army Research Office (ARO) Grant W911NF-14-1-0682; by the ARO Multi-University Research Initiative W911NF-18-1-0218; by the National Science Foundation Grant PHY-1720311; and by the Assistant Secretary of Defense for Research and Engineering via MIT Lincoln Laboratory under Air Force Contract No. FA8721-05-C-0002. A.K. acknowledges support from the NSF Graduate Research Fellowship program. Pacific Northwest National Laboratory is operated by Battelle Memorial Institute under Contract No. DE-AC05-76RL01830 for the U.S. Department of Energy.

%
%

\section*{Author contributions statement}


This research project was a collaboration between experts in quantum systems (A.P.V., A.H.K., F.V., S.G., and W.D.O.) and nuclear physics (J.A.F., J.L.O., B.A.V., B.L., and A.S.D). Simulations of background radiation and the impact of the radiation shielding were performed by A.S.D., B.L., and J.L.O. D.K., A.M., B.N., and J.Y. fabricated the qubit chips. The qubit experiments and data analysis were performed by A.P.V., A.H.K., and F.V. All authors contributed to writing and editing of the Letter.

%
%

\section*{Competing interests}


The authors declare no competing interests.

%
%

%
%

\section*{Data Availability}
The data that support the findings of this study are available from the corresponding author upon reasonable request and with the permission of the US Government sponsors who funded the work.
\section*{Code Availability}
The code used for the analyses is available from the corresponding author upon reasonable request and with the permission of the US Government sponsors who funded the work.

\section*{Additional information}
{\bf Supplementary Information} is available for this paper.\\
\textbf{Correspondence and requests for materials} should be addressed to A.P.V and J.L.O.

%
%

\clearpage

%
%

\section*{Methods}

\subsection*{Measurement setup}

Extended Data Fig. \ref{fig:ex1}a shows the measurement setup used to measure the energy-relaxation times of the qubits. The qubit control pulses are created using a Keysight PXI arbitrary waveform generator. 
%
The in-phase and quadrature pulses are up-converted to the qubit transition frequency using single sideband IQ modulation.
The readout pulses are created similarly. The control and readout pulses are combined and sent to the sample through a single microwave line. There is a total of 60 dB attenuation in the line to reduce the thermal noise from the room temperature and the upper stages of the dilution refrigerator. In the control line there are eccosorb filters before and after the sample, which further reduce the infrared radiation (thermal photons) reaching the qubit. The control line is inductively coupled to readout resonators R1 and R2.

The control pulses are applied to the qubit via 
the readout resonator, which filters the signal at the qubit frequency. 
Nonetheless, by using sufficiently 
large amplitude pulses, the qubits can be excited in \SI{25}{ns}.

The qubit state is determined using dispersive readout via a circuit QED architecture~\citemethods{Wallraff2005}. The dispersive readout is based on the resonator frequency slightly changing depending on the state of the qubit. The change can be detected by using a single measurement tone near the resonator resonance frequency and measuring the transmitted signal in the microwave line. The measurement signal is boosted using a chain of amplifiers. The first amplifier employed is a near-quantum-limited traveling wave parametric amplifier (TWPA), which has a very low noise temperature and gain up to 30 dB. As with all 
parametric amplifiers, the TWPA requires a pump tone, which is driven by a signal generator at room temperature. 
The measurement signal is further amplified by a LNF HEMT amplifier, which is thermally anchored to the 3 K stage of the refrigerator. At room temperature, there is a final pre-amplifier followed by a heterodyne detector. The down-converted in-phase and quadrature intermediate-frequency (IF) signals are digitized using a Keysight PXI digitizer (500 MHz sampling rate) and then further digitally demodulated 
using the digitizer FPGA to extract the measured qubit state. 
Measurement results are ensemble averaged over many such trials to infer the occupation probability of the qubit being in a given state.

In the experiments, we used one sample with 2 transmon qubits, and a second sample with 5 transmon qubits. The qubits were fabricated using optical and electron beam lithography. By construction, the structure of our qubits is kept simple and pristine – MBE-grown aluminum on top of high-resistivity silicon – to reduce defects that cause decoherence. The Josephson junctions have an additional layer of aluminum oxide in between the aluminum leads and are fabricated using double-angle shadow evaporation in another UHV evaporator (different from the MBE). The fabrication is similar to that described in reference~\nocitemethods{yan2016flux} 41.

Extended Data Table \ref{fig:ex1}e shows the relevant qubit parameters. The reported energy relaxation times are median values during the lead-shield experiment. The values for Q1 and Q2 differ from those reported for \ce{^64}{Cu} measurements due to their fluctuation over time.

\subsection*{Production of \ce{^64}{Cu} source}

The $^{64}$Cu radiation source was created by neutron activation of natural copper via the capture process \ce{^63}Cu (n,$\gamma$)\ce{^64}Cu. Given its 12.7 hour half-life, \ce{^64}Cu is well suited for deployment in a dilution refrigerator, since it takes 72--100 hours to cool to base operating temperature. The irradiation took into account the anticipated $^{64}$Cu decay during the cool-down period, by specifically irradiating at higher levels of $^{64}$Cu than used in the qubit study and then allowing the foils time to decay to lower levels of activity.

Two copper disks created from the same McMaster-Carr foil were irradiated with neutrons at the MIT Reactor (MITR).   The two foils are referred to as sample ``A'' and ``A-Ref''. The irradiated sample ``A'' was installed in the dilution refrigerator with the two qubits described in this study, while ``A-Ref'' was kept to determine the level of radioactive activation products.  Each of the foils were \SI{7.5}{\milli\metre} in diameter and $0.5 \pm 0.1$ mm thick and had a mass of 178.5 mg and 177.6 mg, respectively.  The total neutron irradiation exposure was 7~minutes and 14~seconds in duration.  Using a high purity gamma-ray spectrometer, the ``A-Ref'' sample was used to determine the $^{64}$Cu activation level. We determine the activity of sample ``A'' to be $(162 \pm 2)~\mu$Ci at 9:00 AM ET May 13, 2019.  This activity is based on measurements of  $^{64}$Cu's 1346~keV gamma-ray.  

Despite the high copper purity (99.99\%), trace elements with high neutron cross-sections can also be activated from the neutron irradiation process.  The same HPGe counter was used to determine the presence of other trace elements, the results of which are reported in Extended Data Table \ref{fig:ex2}a.

\subsection*{Operation of NaI detector}

A standard commercial NaI detector measures energy deposited in the NaI crystal through the scintillation light created when $\gamma$- or x-rays scatter atomic electrons in the crystal. 
The magnitude of the scintillation light signal, measured by a photomultiplier tube (PMT), is proportional to the energy deposited in the NaI crystal by the incident radiation. 
As the specific energy of $\gamma$- or x-rays are indicative of the radioactively decaying nucleus, an energy spectrum measured by the NaI detector can be used to determine the relative contributions of ionizing radiation in the laboratory due to different naturally occurring radioactive isotopes. 
In a normal laboratory environment, the dominant naturally occurring radioactive nuclei consist of isotopes in the uranium ($^{238}$U) and thorium ($^{232}$Th) decay chains as well as $^{40}$K. 
These features are identified in Fig.~\ref{fig:fig3}a). 


It is possible to reduce the high voltage applied to the PMT, effectively reducing the gain on the scintillation light signal from the NaI detector.
This enables the measurement of ionizing cosmic rays -- and the secondary radiation produced by them -- as determined mainly by spectral features above 2.7 MeV~\citemethods{PhysRevD.98.030001} 
(see Extended Data Fig.~\ref{fig:ex2}e for the measured spectrum). 
We can fit the known spectrum of cosmic rays to the measured spectrum to find the cosmic ray flux in the laboratory. The fit is shown in Extended Data Fig. \ref{fig:ex2}e. Note that below 2.7 MeV the large difference between the measurement and the fit is due to the radiation from nuclear decays, as shown in Fig.~\ref{fig:fig3}a.

\subsection*{Radiation transport simulations and normalization}
To estimate the power density imparted into the qubits by radiation, we developed a radiation transport simulation.  The simulation was performed using the Geant4 toolkit \cite{AGOSTINELLI2003250} which is designed for modeling the interaction of radiation and particles with matter. The simulation geometry included a detailed model of the layers of the Leiden cryogenics CF-CS81 dilution refrigerator, the mounting fixtures and containment for the qubit, and the activated copper foil as it was located for the experiment.  The qubit chip is modelled as a 380 $\mu$m thick piece of silicon with a 200~nm aluminum cladding.  Input power density is estimated by measuring simulated energy deposited into the aluminum layer.  Three separate radiation source terms are considered: \ce{^64}{Cu} and the other isotopes in the activated copper foil, naturally occurring background radiation primarily from the concrete walls of the laboratory, and cosmic ray muons.   

To estimate the effect of isotopes in the copper, we make use of Geant4's radioactive decay simulation capabilities.  Instances of each isotope are distributed uniformly throughout the simulated foil volume.  Geant4 samples the available decay modes for that isotope with appropriate branching fractions, and generates the corresponding secondary particles (gammas, betas, positrons, etc), which are then tracked until they have deposited all their energy.  By tallying up these events, we can estimate the average input energy density into the qubit substrate per decay, or equivalently the average power density per unit of isotope activity. The total simulated spectrum at various times during the qubit measurement campaign are shown in Fig.~\ref{fig:fig3}b. 

To understand the background ionizing radiation levels present in the MIT laboratory where all qubit devices are operated, a $3''\times3''$ NaI scintillator detector was deployed near the dilution refrigerator where the qubit measurements 
were made.  The detector was represented in the radiation transport simulation as a bare NaI cylinder (not including any housing, photomultiplier tube, etc).  Gammas with an energy spectrum following the equilibrium emissions of the most common radioactive isotopes (\ce{^238}{U}, \ce{^232}{Th}, and \ce{^40}{K}) are simulated starting in a sphere surrounding the NaI detector with an isotropic initial direction.  A small number of simulations were run with different-sized initial locations to evaluate the impact of this parameter, yielding a 10\% systematic uncertainty. 

In order to fit to the measured data, the simulated energy deposits must be ``smeared'' to account for the detector's finite energy resolution.  We used a quadratic energy scaling function to map energy to measured ADC counts, and a quadratic resolution function as a function of energy: 
\begin{equation}
    \sigma^2 = \sigma_0^2 + \sigma_1^2E + \sigma_2^2E^2
\end{equation}
Each of the energy scale and resolution coefficients is left free in the fit, as well as the flux of each isotope, for a total of 9 free parameters.  The result for a fit over the range 0.2 - 2.9 MeV is show in Fig.~\ref{fig:fig3}a.  The fit is much better when performed over a narrower region of the data.  This could be improved with a more sophisticated response function, but we address the issue by performing the fit separately over three energy ranges: 0.2-1.3 MeV, 1.3-2.9 MeV, and 0.2-2.9 MeV, and taking the difference as a systematic uncertainty.  This result is reported in the first line of Extended Data Table \ref{fig:ex2}b.  In total the uncertainty in the fits contributes 8\% to the systematic uncertainty.  The simulated energy deposition efficiency for each external isotope is approximately equal to 0.04 keV/s/mm$^3$ per cm$^{-2}$s$^{-1}$, which yields a total power density from environmental gammas of $0.060 \pm 0.005$~keV/s/mm$^3$. 

The same NaI detector, operated at lower gain, is used to estimate the cosmic ray flux, see Extended Data Fig. \ref{fig:ex2}e.  Cosmic ray muons are simulated in a 1~m square plane above the detector, using the CRY package to generate the energy spectrum and angular distribution~\citemethods{CRY}. The muon flux taken directly from CRY is $1.24\times10^{-2} \mathrm{cm}^{-2}\mathrm{s}^{-1}$.  A fit to the low-gain NaI data, using the same convolutional technique as for gammas, yields $(9.7 \pm 0.1) \times10^{-3} \mathrm{cm}^{-2}\mathrm{s}^{-1}$, or about 20\% lower than the CRY value.  The same simulation gives an energy deposition efficiency in the qubits of $(4.3 \pm 0.2)$ keV/s/mm$^3$ per cm$^{-2}$s$^{-1}$ of cosmic ray muon flux.  This, in turn, yields a cosmic-ray induced power density of $(0.042 \pm 0.002)$ keV/s/mm$^3$.

Throughout this work, we have based our analysis on the absorbed power density in the aluminum. 
However, 
radiation will also interact and deposit energy in the silicon substrate. 
How much of this energy, if any, reaches the aluminum layer and is converted to quasiparticles is unknown, in part because we do not know the relevant coupling rates between silicon and aluminum or the various recombination rates of the quasiparticles. 
This motivated our use of a calibrated \ce{^64}{Cu} source, which we use to parameterize the net effect. 
In fact, as we show below, whether we consider aluminum only, or aluminum plus silicon, the difference in the net result changes by at most order unity. 
This somewhat counterintuitive result arises, because the power densities in aluminum and silicon are approximately the same, and because the \ce{^64}{Cu} captures the net effect in either case.

Although \ce{^64}{Cu} captures the net effect well, small differences arise due to how the radiation is emitted and absorbed. For example, in comparison to highly penetrating radiation, \ce{^64}{Cu} deposits a larger fraction of its emitted energy into the aluminum, because a larger fraction is emitted as betas. If the quasiparticle density is dominated by energy from the silicon rather than the aluminum, the relative strength of \ce{^64}{Cu} to the other trace activated isotopes would be approximately 60\% lower.  The external power density induced from environmental gammas is approximately 20\% lower, while the cosmic ray power density is 13\% higher, for a net 7\% total increase in external power.  The lead shielding effectiveness ($\eta$) is also approximately 15\% higher for the silicon than aluminum. By choosing aluminum, we are taking the most conservative estimate for the impact of environmental radiation on qubit energy relaxation.


We now show that these differences are at most an effect of order unity. If, for example, 
the quasiparticle generation rate is dominated by the total absorbed power in the silicon substrate, we can estimate the maximal relative error in the estimate of $\Gamma_1$ by comparing the ratios of the power densities of the external radiation $P_{\rm ext}$ absorbed in the aluminum film and the silicon substrate to the ratios of power densities induced by the \ce{^64}{Cu} source as
\begin{equation}
\label{eq:fc_methods}
    f_{\rm c} = \sqrt{\frac{P^{\ce{Al}}_{\rm src}(t)}{P^{\ce{Si}}_{\rm src}(t)} \times \frac{P^{\ce{Si}}_{\rm ext}}{P^{\ce{Al}}_{\rm ext}}} \approx 1.6.
\end{equation}
This would would increase our estimate of the effect of the external radiation on the qubit energy-relaxation rate from $\Gamma_1^{\rm (Q1)} = 1/(\SI{4}{\milli\second})$ to  $\Gamma_1^{\rm (Q1)} = a\sqrt{\omega_q^{\rm (Q1)} P_{\rm ext}}f_{\rm c} \approx 1/(\SI{2.5}{\milli\second})$. See Supplementary material for the derivation of the above formula. Note that the calculation is an upper-limit estimation, which would be reached only if the total phonon coupling between the silicon substrate and the aluminum is much stronger than the coupling between the sample and the sample holder.

\subsection*{Measurement of the qubit energy relaxation rate}

At the beginning of the measurement, all the qubits are initialized in their ground states. Due to the finite temperature of their environment and hot quasiparticles \cite{Jin2015,Serniak:2018aa}, there is a small excited state population, approximately 1.7\% for these qubits and their qubit transition frequencies. This corresponds to an effective temperature $T_{\rm eff} \approx \SI{40}{\milli\kelvin}$~\cite{Jin2015}. At this temperature, the thermal quasiparticle population can be estimated to be
\begin{equation}
    \label{eq:thermal_qp}
    %
    x_{\rm qp}^{\rm thermal} = \sqrt{2\pi \frac{k_{\rm B} T}{\Delta}}\, {\rm e}^{-\frac{\Delta}{k_{\rm B}T}} \approx \num{7e-24}.
\end{equation}
It is interesting to note that the quasiparticle density $x_{\rm qp} \approx 7\times10^{-9}$ due to ionizing radiation (inferred from our \ce{^64}{Cu}  measurements) would correspond to an equilibrium quasiparticle temperature $T \approx 120$ mK -- consistent with the temperature below which qubit parameters such as $T_1$ stop following an equilibrium quasiparticle model in previous experiments (around 150 mK, see for example References~\citenum{Serniak:2018aa} and~\citenum{Gustavsson1573}).  

The qubit energy relaxation rate $\Gamma_1$ is measured by first driving the qubits to their first excited state using a microwave $\pi$-pulse, see Extended Data Fig. \ref{fig:ex7}b. The state of all the qubits is measured simultaneously after time $t$, which gives an estimate for their residual excited-state population $p(t)$. By changing $t$, the time evolution of the populations can be determined. The model described in Eq.~\eqref{eq:gamma_1} in the main text can be fitted to the measured data to find the energy-relaxation rate $\Gamma_1$ of the qubits.

\subsection*{Estimating the internal radiation rate $P_{\rm int}$}
An accurate estimate of the internal radiation rate $P_{\rm int}$ is important for comparing the feasibility of the shielding effect of the lead shield to the estimated effect of the external radiation power density on the qubit energy-relaxation rate $\delta\Gamma$ extracted from the \ce{^64}{Cu} experiment. A simple way for making the estimate is to extract it from the fit to the data in Fig. \ref{fig:fig3}c. However, the accuracy of the estimate is relatively low, since it is difficult to separate $P_{\rm int}$ from the energy-relaxation rate of the qubit due to sources other than quasiparticles, $\Gamma_{\rm other}$. 
In principle, it is possible to distinguish the two sources, because according to Eq. \eqref{eq:simple_qp_model}, the scaling of $\Gamma_1$ is proportional to $\sqrt{P_{\rm ext} + P_{\rm int} + P_{\rm src}}$ whereas the internal energy-relaxation rate $\Gamma_{\rm other}$ contributes linearly to $\Gamma_1$, see Eq. \eqref{eq:gamma_1}. In practice, this is quite inaccurate, especially if quasiparticle loss is not the dominating loss-mechanism. 

Instead, we employ the shielding experiment to calculate an upper bound for $P_{\rm int}$. In the limit of $P_{\rm int} \gg P_{\rm ext}$, we can calculate an asymmetry parameter for the energy-relaxation times in the shield up or down positions,
\begin{equation}
\label{eq:asymmetry}
    A_i = 2\frac{\Gamma_1^{{\rm d}, i} - \Gamma_1^{{\rm u}, i}}{\Gamma_1^{{\rm d}, i} + \Gamma_1^{{\rm u}, i}} \approx \frac{\eta^{\rm u} - \eta^{\rm d}}{2}\frac{P_{\rm ext}}{P_{\rm int} + \frac{\Gamma_{\rm other}}{a\sqrt{\omega_{\rm q}}}},
\end{equation}
where the index $i$ refers to different rounds of the shield up/down experiment. The internal radiation rate $P_{\rm int}$ can be estimated using the experimentally measured median asymmetry parameter as
\begin{equation}
\tilde{P}_{\rm int} \approx  \frac{(\eta^{\rm u} - \eta^{\rm d})}{2\langle A \rangle}P_{\rm ext} = \SI{7.9}{\kilo\electronvolt\per\cubic\milli\metre\per\second},
\end{equation}
where $\tilde{P}_{\rm int} = P_{\rm int} + \frac{\Gamma_{\rm other}}{a\sqrt{\omega_q}}$, and $\langle A \rangle \approx 0.0028$ (see Extended Data Fig. \ref{fig:ex8}). This gives the upper bound for $P_{\rm int}$. Due to the other relaxation mechanisms, the actual value of $P_{\rm int}$ is lower. For example, $\Gamma_{\rm other} = \SI{1/200}{\per\micro\second}$ would yield $P_{\rm int} \approx \SI{1.6}{\kilo\electronvolt\per\cubic\milli\metre\per\second}$ for the parameters of Q1.
Here we emphasize that the estimate of the asymmetry parameter is based on the data gathered on all the seven qubits employed in the lead shield experiment, with all the qubits having different (fluctuating) values of $\Gamma_{\rm other}$.

\subsection*{Efficiency of the lead shield}
The reduction factor of external $\gamma$-radiation by the lead shield was evaluated using the radiation transport simulation described previously. In the simulation, $\gamma$-rays with energies drawn from the equilibrium emission spectra for \ce{^238}{U}, \ce{^232}{Th}, and \ce{^40}{K} were generated isotropically from the surface of a sphere with 2.4~m diameter, centered on the qubits. The sphere comletely enclosed the model for the lowered lead shield and the dilution refrigerator. The fraction of flux $\Phi$ reaching a smaller 17~cm diameter sphere (fully inside the DR) was recorded.  Extended Data Table \ref{fig:ex2}b shows the results for the no shield, shield down, and shield up, as well as the individual shield efficiency values $\eta^{\rm u} = 1 - (\Phi^{\rm u}/\Phi^{\textrm{no shield}})$ and $\eta^{\rm d} = 1 - (\Phi^{\rm d}/\Phi^{\textrm{no shield}})$.

 A similar simulation was performed to calculate the efficiency of the lead shield against cosmic rays. As expected, the lead shield is ineffective at blocking cosmic rays, but works well against $\gamma$-rays originating from the nuclear decay events in the laboratory, see Extended Data Table \ref{fig:ex2}c.

To validate the simulations, the NaI detector was operated separately inside the lead shield at the approximate location of the qubits in the shield-up configuration.  This configuration was also simulated, and the output fit to the measured spectrum using the same fit procedure as for the bare NaI.  If the simulation and fit procedure are accurate, both fits should give the same values for the input flux.  The results are reported in the first rows of Extended Data Table \ref{fig:ex2}b. The results for U and Th are  consistent, while the values for K differ by about 2.5$\sigma$. It may be that the lead itself has a high level of \ce{^40}{K}, but we treat this as a systematic uncertainty, which is $7\%$ of the total gamma flux. 

Extended Data Fig. \ref{fig:ex1}b-d shows a diagram of the lead shield and its dimensions.

\subsection*{Statistical analysis of the lead shield experiment}
Since there are significant fluctuations in the internal energy relaxation rates $\Gamma_{\rm other}$ of the qubits,  we performed a careful A/B test to verify that the effect of the lead shield on the qubit energy-relaxation time was not due to statistical error. In the measurement of the energy relaxation rates of the qubits, there is uncertainty both due to the measurement accuracy and the fluctuations and drifts in the energy relaxation rates over time. To reduce the uncertainty due to the measurement accuracy, we measured the energy relaxation rates $N$ times in each step of the A/B test. After $N$ measurements the position of the lead shield was swapped (up versus down) and we performed another $N$ measurements. This cycle was repeated 65 times with a sample containing qubits Q1 and Q2. To accelerate data acquisition, 
we installed a second sample with 5 qubits (Q3-Q7) and repeated the measurement cycle an additional 85 times. 
We used $N=50$ for qubits Q1 and Q2, and $N=10$ for qubits Q3-Q7, see Extended Data Fig.  \ref{fig:ex7}a and c for the measured energy-relaxation rates. The median energy-relaxation times of the qubits are listed in the Extended Data Table \ref{fig:ex1}e.

In the spirit of a Dicke radiometer experiment, performing repeated short measurement cycles was crucial for reducing the uncertainty in the relaxation rates due to drifts that occurred on time scales longer than the cycle period.
The drift has been attributed in part to fluctuating two-level systems in dielectrics close to the qubit and in the junction region. 
However, by raising and lowering the shield often enough (every 50th measurement for qubits Q1 and Q2, and every 10th measurement for qubits Q3-Q7) the slow drift is mostly cancelled. 
Extended Data Fig. \ref{fig:ex7}d shows the spectral density of the $T_1$ noise for qubits Q3, Q4, Q6, and Q7. The noise power density approximately follows a power law $S = {\rm const}/f^\alpha$ with $\alpha \approx{1.5}$. The fit to the model is shown with solid orange line. The noise power density at the lead shield up/down cycle frequency of 1/($15 \pm 1$\si{\min}) is $\SI{3.4e4}{\micro\second\squared\per \hertz}$. The noise power in the measurement can be estimated by integrating the spectral density over the noise bandwidth, which for the lock-in measurement yields \SI{49}{\micro\second\squared} (shaded red area in 
Extended Data Fig. \ref{fig:ex7}d). If all the data was gathered sequentially, the noise power can be estimated to be \SI{718}{\micro\second\squared} (gray shaded are in Extended Data Fig. \ref{fig:ex7}d), over an order of magnitude higher than in the Dicke experiment.

We used the median to estimate the net change of $\delta\Gamma_{1}$ (between shield-up and shield-down configurations) to reduce sensitivity to individual measurement outliers. The quasiparticle contribution to the energy relaxation rates of the qubits depends on their frequencies according to Eq.~\eqref{eq:gamma_power}, and therefore we have normalized the changes in the energy relaxation rates to the frequency of Q1 by multiplying by a conversion factor $\sqrt{\omega_{\rm q}^{(Q1)}/\omega_{\rm q}^{(Qi)}}$.

We neglected a small percentage of the total data points where $\Gamma_1^{\rm u}$ or $\Gamma_1^{\rm d}$ were less than \SI{1/30}{\per\micro\second} or their difference was more than 10 standard deviations of all the measured differences, as these tended to indicate suspect rates derived from poorly resolved decay functions. 
%
%
We then calculated the 95\% confidence intervals for $\delta\Gamma_{1}$ using the normal approximation for the confidence interval of the sample median~\citemethods{mangiafico2016summary}.

We applied the Wilcoxon signed-rank test to determine if the median of the two 
distributions (corresponding to the shield-up versus shield-down configurations) differ in a statistically significant manner. 
This is a non-parametric test and can be used for data that are not normally distributed. 
For $\delta\Gamma_{1}$, the single-sided Wilcoxon signed-rank test gives a p-value $p= 0.006$ for the null hypothesis that the median of the energy relaxation rates with the shield is the same or higher than without the shield. The test statistic $w \approx \num{25000000}$ with a sample size of 9846.
For a p-value $\ll 0.05$, we can reject this null hypothesis and conclude that the shield reduces the energy-relaxation rate.

We performed several tests to verify the correctness of our statistical analysis. First, we checked that the result is not sensitive to the post-processing we performed on the data. The first panel of Extended Data Fig. \ref{fig:ex6}a shows the p-value of the Wilcoxon signed-rank test for a range of different cutoff parameters. The p-value remains low for all the sensible parameters we tested, verifying that the finding is not an artifact of post-processing or parameter selection. The median value is even less sensitive to the post-processing, shown in the lower-left panel. The blue diamond in the upper-left corner shows the point where no post-processing is done. The blue circle shows the values which we use in the main text, $T_1^{\rm cutoff} = \SI{30}{\micro\second}$ and $n^{\rm cutoff}_\sigma = 10$.  

Next, we tried shuffling the data by comparing the energy relaxation rates of the measurements to the next measurement without moving the shield. In this case, we would expect the signal to completely vanish, and the null hypothesis to be manifestly true. The result is shown in the middle column of Extended Data Fig. \ref{fig:ex6}a. In this case, the p-value is close to 1, which implies that we must accept the null-hypothesis that there is no signal if we don't move the shield, as expected.

In the third test, we completely randomized the pairs of measurements which we compare, resulting in overall high p-value, supporting our analysis (third column).

Extended Data Fig. \ref{fig:ex6}b shows a cutoff of Extended Data Fig. \ref{fig:ex6}a along the dashed lines in the left and middle panels. The filled areas show the 95 \% confidence interval of the medians.
\bibliographystylemethods{naturemag-doi}
\bibliographymethods{references}

\clearpage
\renewcommand\thefigure{\arabic{figure}}
\renewcommand{\figurename}{Extended Data Fig.}
\setcounter{figure}{0}    
\begin{figure*}[htb]
    \centering
    \includegraphics[width=157mm]{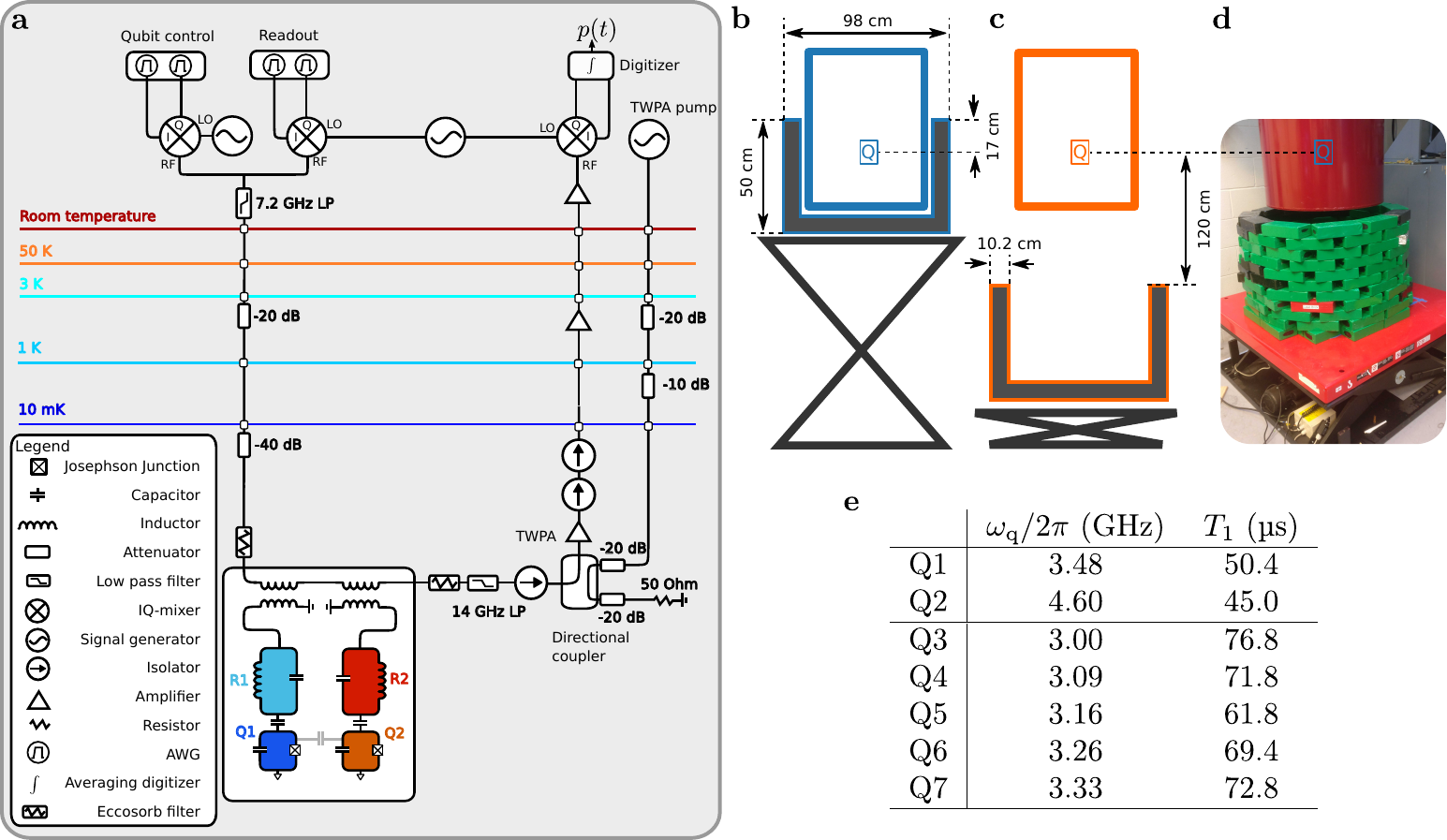}
    \caption{{\bf Experimental set-up.} {\bf a} The diagram shows a simplified block diagram of the room temperature electronics and dilution refrigerator configuration used for measuring the qubit frequency and coherence times. Panels {\bf b-d} show the schematic of the lead shield used to block environmental radiation. The lead shield can be raised and lowered using a scissor lift. {\bf b} In the up-position, the qubits were \SI{17}{\centi\metre} below the edge of the lead shield. {\bf c} In the lowered position the edge of the lead shield was \SI{120}{\centi\metre} below the qubits. {\bf d} Picture of a partially-raised lead shield [in between the configurations shown in panels {\bf b} and {\bf c}]. The lead bricks are wrapped in protective plastic film. {\bf e} The parameters of the qubits used in the lead shield experiment.}
    \label{fig:ex1}
\end{figure*}

\clearpage

\begin{figure*}[t]
    \centering
    \includegraphics[width=89mm]{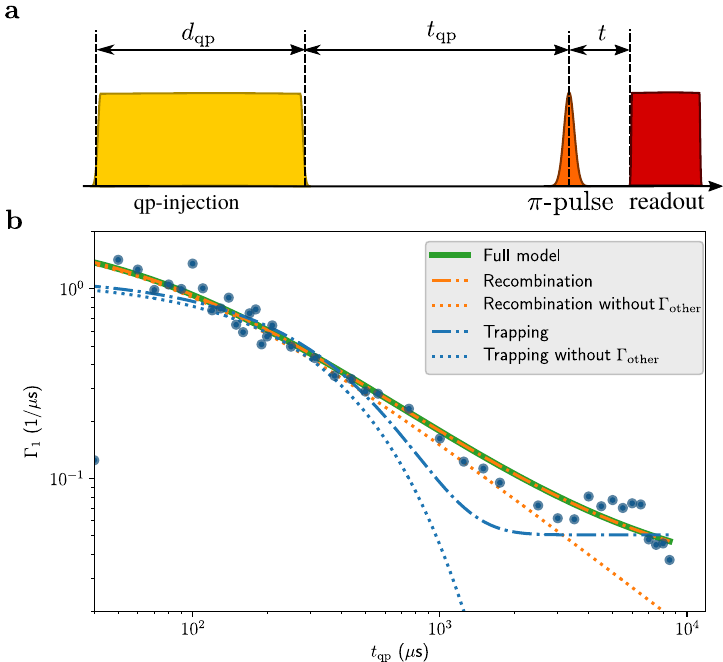}
    \caption{{\bf Quasiparticle injection experiment.} {\bf a} shows the pulse sequence in the quasiparticle injection experiment. First a strong microwave pulse is applied for the duration of $d_{\rm qp}$ to the resonator, which excites quasiparticles. After time $t_{\rm qp}$ the { energy-relaxation time} of the qubit is measured. {\bf b} shows the energy relaxation rate of the qubit Q1 during the quasiparticle injection experiment (blue dots). A solid green line shows a fit to the data using the full model that includes quasiparticle trapping and recombination. Orange dash dotted line shows the model with only recombination, dotted line shows the same model without the internal quasiparticle relaxation rate $\Gamma_{\rm other}$. Blue dash dotted line shows the fit to the model that only includes trapping of quasiparticles. Dotted blue line shows the trapping model without $\Gamma_{\rm other}$.}
    \label{fig:ex3}
\end{figure*}

\clearpage

\begin{figure*}[t]
    \centering
    \includegraphics[width=157mm]{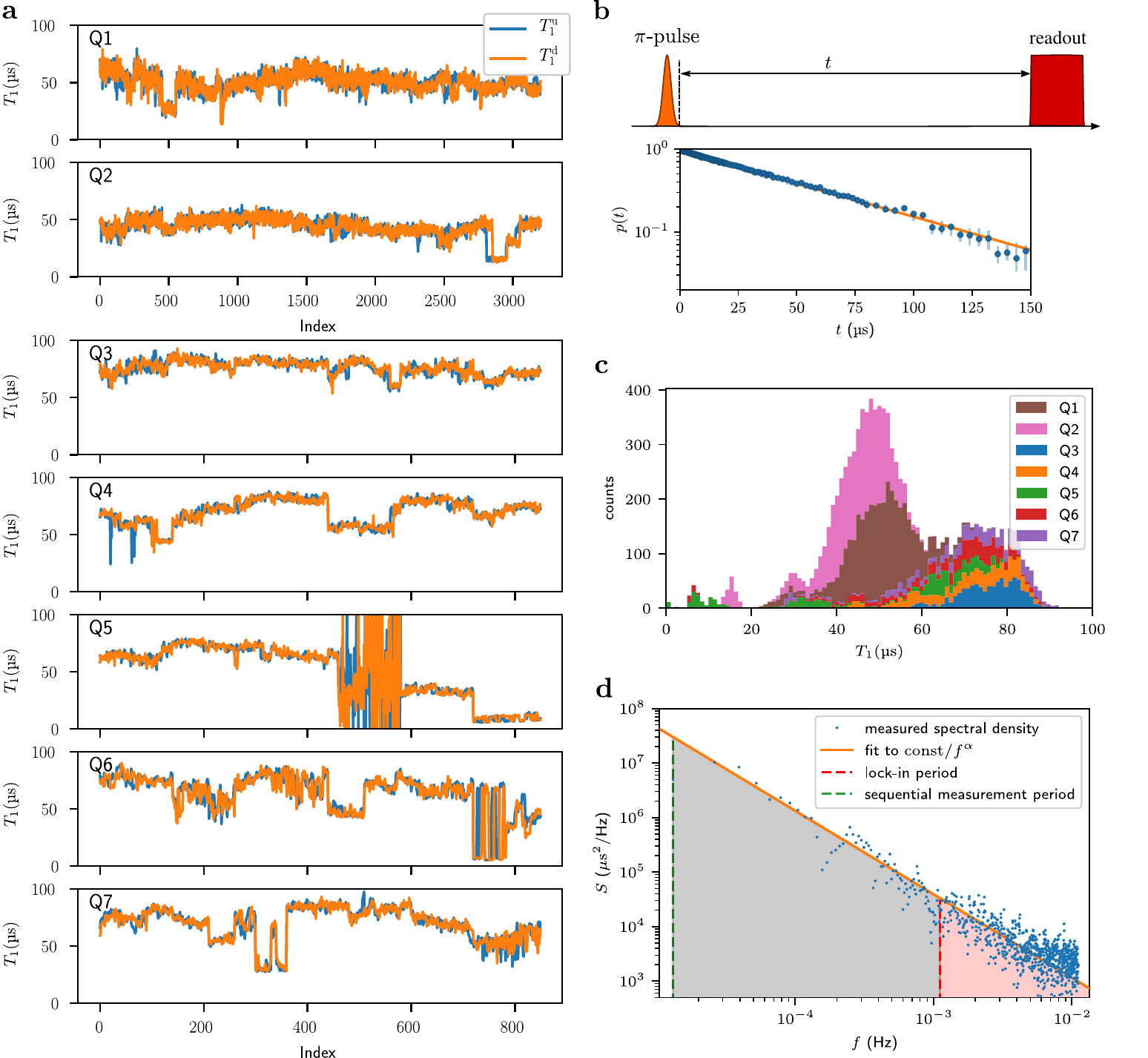}
    \caption{{\bf Energy-relaxation times in the shielding experiment}. Energy relaxation times $T_1$ for qubits Q1-Q7 during the lead shield experiment while the shield is in up (blue) or down (orange) positions. {\bf b} shows the pulse sequence used to measure the energy relaxation rate of all the qubits. First a $\pi$ pulse is applied to all the qubits. After time $t$ a measurement pulse is used to determine the state of the qubits. The qubit excited state population relaxes exponentially as a function of time. Blue circles show the measured qubit excited state populations and the orange line is an exponential fit using the model of Eq. \eqref{eq:gamma_1}. {\bf c} Stacked histogram of the combined energy relaxation times for all of the qubits in the lead shield experiment. {\bf d} Plot of the noise power spectral density during the lead shield experiment for qubits Q3, Q4, Q6, and Q7. The red dashed red line marks the rate of a single cycle of the lead shield. The green dashed line shows the estimated measurement period if all the data was gathered sequentially. Orange line is a fit to a power law $S = {\rm const}/f^\alpha$ with $\alpha \approx 1.5$.}
    \label{fig:ex7}
\end{figure*}

\clearpage

\begin{figure*}[t]
    \centering
    \includegraphics[width=89mm]{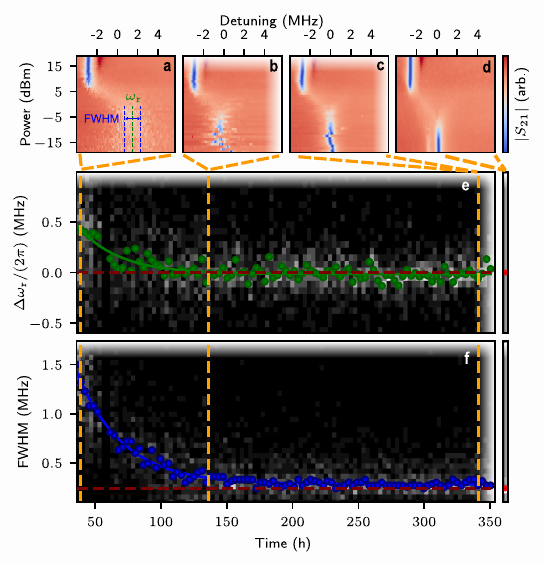}
    \caption{\textbf{Resonator single-tone spectroscopy.} \textbf{a-d} show the transmission profile of resonator 1 as as a function of readout power and readout frequency at different times throughout the experiment. When exposed to a high level of radiation, the resonator frequency becomes unstable in the dispersive regime which is used for reading out the qubit. The resonator becomes more stable as the radiation source decays. \textbf{e} shows the change in the resonance frequency, $\Delta \omega_{\text{r}}$, due to radiation through out the experiment. We observe that the median $\Delta \omega_{\text{r}}$ follows an exponential decay with a half-life of $t_{1/2}= (21.74 \pm 2.8) \si{\hour}$. Furthermore, in \textbf{f} we see the full-width-half-max (FWHM) of the resonator also exponentially decay with a half-life of $t_{1/2}= (24.16 \pm 0.78) \si{\hour}$ until converging to the control value.}
    \label{fig:ex4}
\end{figure*}
\clearpage

\begin{figure*}[t]
    \centering
    \includegraphics[width=89mm]{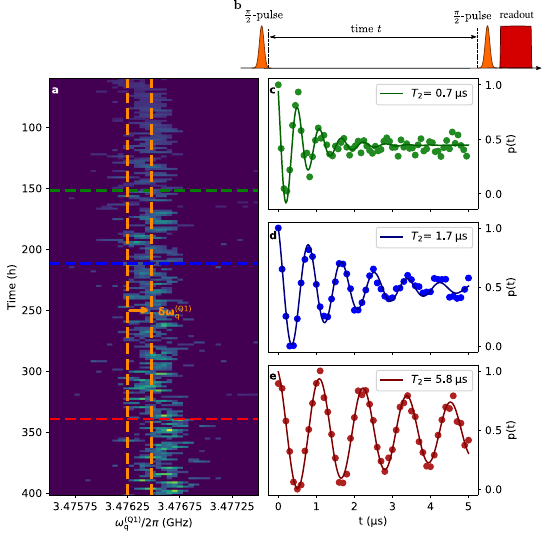}
    \caption{{\bf Qubit frequency shift.} {\bf a} The frequency of the qubit can be determined from a Fourier transform of a Ramsey measurement, shown in the panel at different times after installation of the \ce{^64}{Cu} source. We plot the inferred qubit frequency by offsetting the measured Fourier transform spectra by the frequency of the control pulses. The orange dashed lines show the shift in the average qubit frequency during the experiment. {\bf b} The pulse sequence used in a Ramsey measurement. First $\pi/2$-pulse prepares the qubit in a superposition state. The phase of the qubit state evoles during time $t$, after which a second $\pi/2$-pulse is applied before the measurement pulse. {\bf c-e} Ramsey oscillations and fit $T_2$ times are shown at \SI{152}{\hour}, \SI{212}{\hour} and \SI{340}{\hour} after installation of the \ce{^64}{Cu} source. The dashed lines in panel {\bf a} show the times at which the measurements are performed.}
    \label{fig:ex5}
\end{figure*}

\clearpage

\begin{figure*}[t]
    \centering
    \includegraphics[width=157mm]{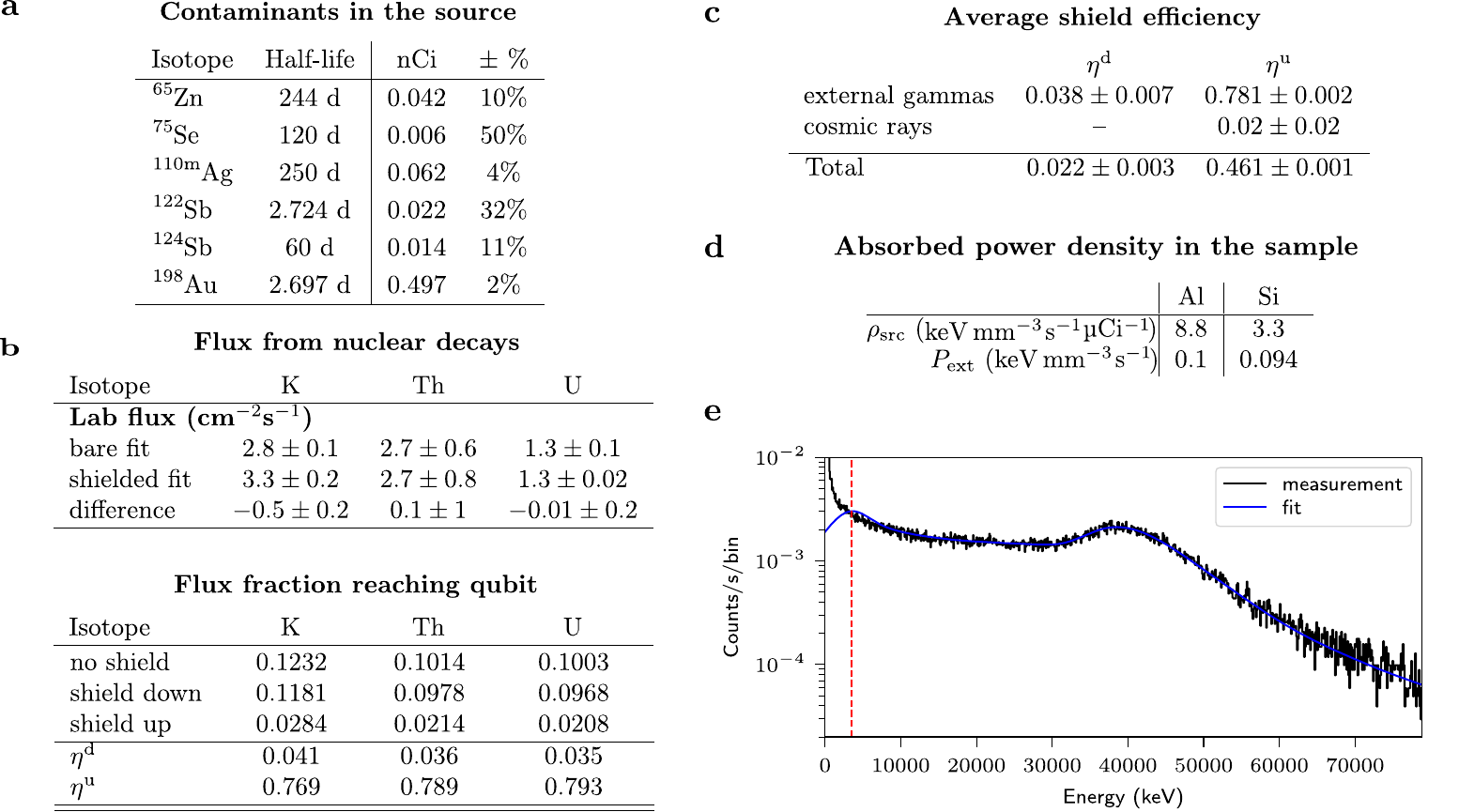}
    \caption{{\bf Radiation transport simulations.} {\bf a} Isotopes measured to be present in the sample ``A-Ref'' and the activity inferred for each in sample ``A'' as of May 24, 2019 at 4:00 PM Eastern time zone. {\bf b} Results from simulations of environmental radiation sources in the laboratory environment.  The background gamma flux is obtained by a fit to a measurement with a NaI scintillator (Fig.~\protect{\ref{fig:fig3}}a), simulating and measuring both with and without the lead shield in the ``up'' position. Cosmic rays were also measured and simulated for both shield-up and shield-down conditions; the shield did not have a measurable effect in the ``up'' position, as expected, and the effect is taken to be zero in the ``down'' position. {\bf c} The Average shield efficiency values for $\eta$ are weighted by each component's contribution to total external power. 
    Statistical uncertainties on the fraction of flux reaching the interior of the DR are all 0.0001; uncertainties on $\eta$ values for individual isotopes are all approximately 0.001. {\bf d} Power densities absorbed in silicon and aluminum. {\bf e} The figure shows the spectrum of the energy deposited in a NaI detector by cosmic ray muon secondaries measured in the laboratory. Blue solid line shows the known cosmic ray muon spectrum fit to the measured data. The spectrum corresponding to energies below the dashed red line is shown in Fig. \ref{fig:fig3}a. Note that in the spectrum shown here a different energy bin width is used to capture higher energy scales.}
    \label{fig:ex2}
\end{figure*}

\clearpage

\begin{figure*}[t]
    \centering
    \includegraphics[width=157mm]{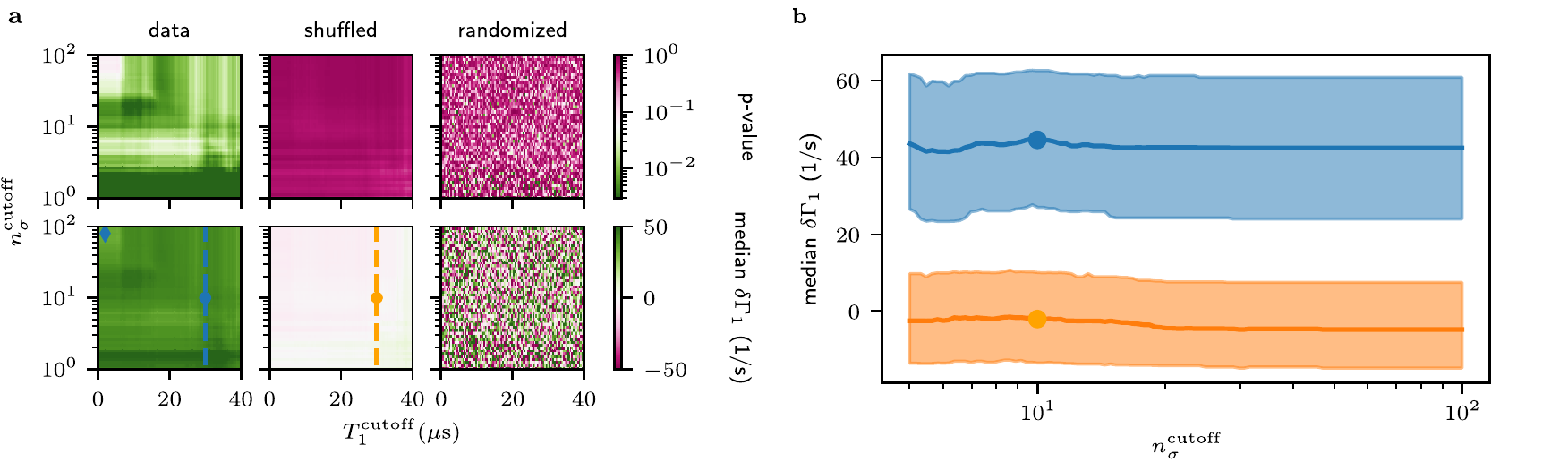}
    \caption{{\bf The effect of post-processing on the lead shield effect A/B test.} {\bf a} The upper row shows the p-value of the Wilcoxon signed rank test for three different test cases and for the different post-processing parameters. On the horizontal axis the $T_1^{\rm cutoff}$ is varied. The vertical axis shows the effect of applying a cutoff to the difference in the energy relaxation rates when the shield status is changed. The first column shows the actual data. The middle column shows a reference experiment, where the energy relaxation rates are compared without moving the shield. The last column shows the data when the energy relaxation rate pairs are randomized. The lower row shows the median of the effect of the shield on the energy relaxation rate $\delta\Gamma_1$. {\bf b} The median of $\delta\Gamma_1$ along the dashed lines in {\bf a}. The filled area shows 68\% confidence intervals for the median.}
    \label{fig:ex6}
\end{figure*}

\clearpage

\begin{figure*}[t]
    \centering
    \includegraphics[width=89mm]{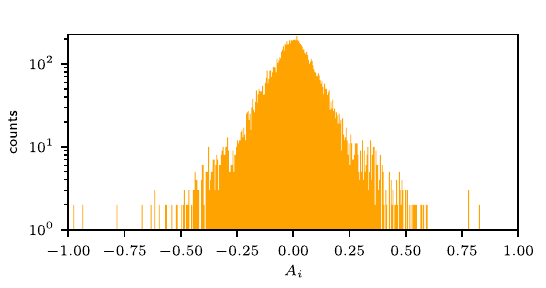}
    \caption{{\bf Asymmetry parameter distribution.} The distribution of the asymmetry parameter $\langle A \rangle$ of the energy relaxation rates between the shield in up or down position.}
    \label{fig:ex8}
\end{figure*}

\newpage
\clearpage

\section*{Supplementary material}

\subsection*{Quasiparticle injection experiment}

Quasiparticles can be injected into the circuit to study their relaxation dynamics. Here we attempt to determine whether the quasiparticle dynamics are dominated by the recombination rate $r$ or trapping rate $s$, see Eq. \eqref{eq:simple_qp_model} in the main text. In the absence of the external generation rate, the time evolution of the quasiparticle density can be solved from the differential equation in Eq.~\eqref{eq:simple_qp_model}, see also Reference~\citenum{Wang2014} in the main text,

\begin{equation}
\label{eq:full_qp_evolution}
x_{\rm qp}(t) = \frac{x_{\rm qp}(0) s}{-r x_{\rm qp}(0) + \mathrm{e}^{s t}(s + r x_{\rm qp}(0))}.
\end{equation}
In the two limiting cases of no trapping or no recombination the time evolution can be simplified as

\begin{equation}
\label{eq:recombination_qp_evolution}
x_{\rm qp}(t) = \frac{x_{\rm qp}(0)}{1 + x_{\rm qp}(0) r t}
\end{equation}
or

\begin{equation}
\label{eq:trapping_qp_evolution}
    x_{\rm qp}(t) = x_{\rm qp}(0)\mathrm{e}^{-s t},
\end{equation}
respectively.

Following the experimental protocol introduced in Reference~\citenum{Wang2014} in the main text, quasiparticles can be generated by strongly driving the resonator coupled to Q1. The energy pumped into the resonator creates a voltage over the Josephson junction of the qubit and breaks Cooper pairs, resulting in an elevated quasiparticle density. The generated quasiparticles then gradually diffuse into the superconducting material around the Jospehson junction. We observed that a steady state in the quasiparticle density in the qubit was reached after $d_{\rm qp} = \SI{10}{\milli\second}$ of quasiparticle injection, see Fig. \ref{fig:ex3}a for the pulse sequence. After the initial quasiparticle injection pulse, the quasiparticle density was estimated by measuring the qubit energy relaxation rate, see Eq.~\eqref{eq:gamma_qp} in the main text. By changing the delay between the injection pulse and the energy relaxation rate measurement, we can determine the quasiparticle relaxation dynamics in our device, see Extended Data Fig.~\ref{fig:ex3}. We fitted the full model of Eq.~\eqref{eq:full_qp_evolution}, shown by the solid green line in Extended Data Fig.~\ref{fig:ex3}b.
The fit includes the internal relaxation rate of the qubit, $\Gamma_{\rm other} \approx \SI{1/35}{\per\micro\second}$. The dash dotted orange line shows the fit using a model that only includes recombination. This line almost exactly matches the fit using the full model, confirming our assumption that recombination is the dominant quasiparticle decay process in our devices. The dotted orange line assumes no internal energy relaxation $\Gamma_{\rm other} = 0$. The blue dash dotted line shows the model which assumes trapping as the only decay mechanism of quasiparticles. The dotted blue line shows the model without $\Gamma_{\rm other}$. The model that assumes only trapping is strongly disfavored by the data.
From the full model, we extract values for $s$ and $r$ and conclude that the action of the quasiparticle trapping rate $s$ is negligible compared with the recombination rate $r$.

\subsection*{Resonator measurements}

Each qubit on the device is addressed and operated via a separate resonator using a microwave probe pulse in the qubit-resonator dispersive regime. 
To experimentally determine the resonant frequency of the resonator, we scan both the probe frequency and probe power and record the response. 
In the absence of radiation, at relatively low powers, the resonator frequency depends on the qubit state and serves as the basis for dispersive qubit readout. 
In the high-power regime, the resonator effectively decouples from the qubit and its frequency is ``locked'' and (ideally) independent of the qubit state. 
We systematically repeat this frequency and power scan at different source radiation intensities to study the behavior of our resonators in the presence of ionizing radiation.
In the presence of radiation, in principle, such a measurement could provide a means to distinguish the impact of radiation on the qubit-resonator system and on the resonator itself, although we have not yet investigated this aspect in sufficient detail to comment on it further here. 

When exposed to ionizing radiation, the resonators become unstable and exhibit random fluctuations in their resonance frequency. As the radiation intensity decreases with time, the fluctuation decreases until the resonator is stable once again (Fig. \ref{fig:ex4}\textbf{a-d}). This behavior is consistent with previous measurements of superconducting resonators in the presence of quasiparticles \citesupp{DeVisser2014,Goldie2013}. We monitor the resonator frequency and full-width-half-max (FWHM) throughout the duration of the \ce{^64}{Cu} radiation experiment. We observe that as the radiation power decreases, both properties of the resonator fluctuations decrease until they converge to the value measured during our control experiment. Furthermore, the median of the resonator frequency shift ($\Delta \omega_\text{r}$) and the FWHM for our measurements follows an exponential decay as a function of time with a half-life of $t_{1/2}= (21.74 \pm 2.8) \si{\hour}$ and $t_{1/2}= (24.16 \pm 0.78) \si{\hour}$ respectively. The observed decay half-life values are very close to being twice the half-life of the \ce{^64}{Cu} source. This effect can be explained by quasiparticle induced change in the kinetic inductance of the resonator. The kinetic inductance of superconducting resonators is directly correlated with the number of quasiparticles \citesupp{Barends2008}: $\delta L_k/L_k=\frac{1}{2} \delta n_{\text{qp}}/ n_{\text{qp}}$. Furthermore, the change in the resonator frequency is directly proportional to the change in the kinetic inductance: $\delta\omega/\omega_0 = \frac{\alpha}{2} \delta L_k/L_k$. Therefore, $\delta\omega/\omega_0 \propto \delta n_{\text{qp}}/ n_{\text{qp}}$. According to Eq. \eqref{eq:gamma_power} in the main text, the quasiparticle density depends on the square root of radiation power, and therefore we expect the resonator frequency decay constant to be twice that of the radiation source.

\subsection*{Absorption of radiation power in the sample}
According to Eq. \eqref{eq:gamma_qp}, the energy relaxation rate due to ionizing radiation power density absorbed in the aluminum comprising the qubit can be written as
\begin{equation}
\label{eq:eq0}
\Gamma_1(t) = a\sqrt{\omega_{01} P_{\rm src}^{\rm Al}(t)},
\end{equation}
where $P_{\rm src}$ is the power density of the \ce{^64}{Cu} source in aluminum. Here we calculate how much bigger the impact of ionizing radiation is on the energy-relaxation rate of the qubit if we consider total absorbed power in the entire sample -- aluminum plus silicon -- instead of the absorbed power in aluminum only and, ultimately, the absorbed power density in aluminum, as considered in the main text. As we will show, the use of power density in aluminum differs from the total absorbed power by a factor of order unity for our experimental conditions (absorbed power densities in aluminum and silicon are similar in our experiment). 

The total power absorbed in aluminum is $V_{\rm Al} P_{\rm src}^{\rm Al}(t)$, and we can write
\begin{equation}
\label{eq:eq1}
\Gamma_1(t) = \alpha\sqrt{\omega_{01} V_{\rm Al} P_{\rm src}^{\rm Al}(t)},
\end{equation}
where $V_{\rm Al}$ is the aluminum volume, and $\alpha$ is an unknown coefficient. Alternatively, one can consider the total power absorbed in the whole sample,
\begin{equation}
\label{eq:eq2}
\Gamma_1(t) = \beta\sqrt{\omega_{01} \left(V_{\rm Al} P_{\rm src}^{\rm Al}(t) + V_{\rm Si} P_{\rm src}^{\rm Si}(t)\right)},
\end{equation}
where $V$ are the volumes of the materials and $\beta$ is a different constant. The energy-relaxation rate of the qubit due to external radiation can be similarly written as

\begin{equation}
\Gamma_{\rm ext}^{\alpha} = \alpha\sqrt{\omega_{01} V_{\rm Al}P_{\rm ext}^{\rm Al}},
\end{equation}
for the absorption in aluminum, or as 
\begin{equation}
\Gamma_{\rm ext}^\beta = \beta\sqrt{\omega_{01} \left(V_{\rm Al} P_{\rm ext}^{\rm Al} + V_{\rm Si} P_{\rm ext}^{\rm Si}\right)}.
\end{equation}
We can compare the difference of these two ways of estimating $\Gamma_{\rm ext}$ by solving for their relation
\begin{equation}
    \label{eq:fc}
    f_{\rm c} = \frac{\Gamma_{\rm ext}^\beta}{\Gamma_{\rm ext}^\alpha} = \frac{\beta}{\alpha} \sqrt{\frac{V_{\rm Al} P_{\rm ext}^{\rm Al} + V_{\rm Si} P_{\rm ext}^{\rm Si}}{V_{\rm Al}P_{\rm ext}^{\rm Al}}}.
\end{equation}
We can solve for $\alpha$ and $\beta$ in Eqs. \eqref{eq:eq1} and \eqref{eq:eq2} and substitute those to Eq. \eqref{eq:fc} to yield

\begin{equation}
\label{eq:fc_subst}
f_{\rm c} = \sqrt{\frac{V_{\rm Al}P_{\rm src}^{\rm Al}(t)}{V_{\rm Al} P_{\rm src}^{\rm Al}(t) + V_{\rm Si} P_{\rm src}^{\rm Si}(t)}}\sqrt{\frac{V_{\rm Al} P_{\rm ext}^{\rm Al} + V_{\rm Si} P_{\rm ext}^{\rm Si}}{V_{\rm Al}P_{\rm ext}^{\rm Al}}}.
\end{equation}
Assuming $V_{\rm Al} \ll V_{\rm Si}$ and that the absorbed power densities in all the materials are similar, we can simplify the above equation to give
\begin{equation}
\label{eq:fc_subst_2}
f_{\rm c} \approx \sqrt{\frac{V_{\rm Al}P_{\rm src}^{\rm Al}(t)}{V_{\rm Si} P_{\rm src}^{\rm Si}(t)}}\sqrt{\frac{V_{\rm Si} P_{\rm ext}^{\rm Si}}{V_{\rm Al}P_{\rm ext}^{\rm Al}}},
\end{equation}
and thus
\begin{equation}
\label{eq:fc_final}
f_{\rm c} \approx \sqrt{\frac{P_{\rm src}^{\rm Al}(t)}{P_{\rm src}^{\rm Si}(t)}}\sqrt{\frac{P_{\rm ext}^{\rm Si}}{P_{\rm ext}^{\rm Al}}},
\end{equation}
which is the same as Eq. \eqref{eq:fc_methods} in the methods section. The absorbed power density from the \ce{^64}{Cu} source is given by $P_{\rm src}(t) = A(t)\rho_{\rm src}$, where $A(t)$ is the activity of the source as a function of time. The power densities are provided in the Extended Data Table \ref{fig:ex2}d. Finally, the correction coefficient is
\begin{equation}
    \label{eq:fc_final_2}
f_{\rm c} \approx \sqrt{\frac{\rho_{\rm src}^{\rm Al}}{\rho_{\rm src}^{\rm Si}}}\sqrt{\frac{P_{\rm ext}^{\rm Si}}{P_{\rm ext}^{\rm Al}}} \approx 1.6,
\end{equation}
which is of order unity.
Note that the result does not depend on the volumes of either aluminum or silicon, and therefore the above error estimate can be calculated either using the total absorbed power in the materials, the power densities in the materials, or any combination of the two, as long as the power densities are similar and the assumption $V_{\rm Al} \ll V_{\rm Si}$ holds. Even though the impact in our estimate for $\Gamma_{\rm ext}$ does not significantly depend whether we consider power density in the aluminum or the total absorbed power in the sample, the numeric value of the coefficient $a$ would be scaled by the sample volume if we considered the total absorbed power. Therefore, if the total absorbed power dominates the quasiparticle generation, changing the volume of the sample would have an impact on the quasiparticle density as opposed to the approximation where we only consider the absorbed power density near the qubit. Verifying which of these models better describe our superconducting samples is a topic for future research.

\subsection*{$T_2$ and qubit frequency shift due to ionizing radiation}
{ During the radiation exposure measurement, the $T_2$ coherence times of the qubits were measured in addition to the energy-relaxation rate. The quasiparticles are expected to shift the qubit transition frequency $\delta\omega_{\rm q}$ as

\begin{equation}
\delta\omega_{\rm q} = - \sqrt{\frac{\Delta\omega_{\rm q}}{2\hslash\pi^2}} x_{\rm qp},
    \label{eq:qubit_freq_xqp}
\end{equation}
where $x_{\rm qp}$ is the fraction of broken Cooper pairs and $\Delta$ is the superconductor gap \citesupp{catelani2011}.

In addition to an average shift in the qubit frequency, the coherence time $T_2$ of the qubit is 
reduced due to frequency fluctuations caused by the fluctuating quasiparticle density.
Extended Data Fig.~\ref{fig:ex5}a shows the qubit frequency as a function of time during the exposure to radiation from \ce{^64}{Cu} source measured using a Ramsey measurement, see Extended Data Fig. \ref{fig:ex5}b. Panel a shows the Fourier transform of the measured Ramsey oscillations during the \ce{^64}{Cu} exposure experiment. The frequency of the oscillations corresponds to the difference between the qubit transition frequency and the frequency of the control pulses, $\omega_{\rm R} = \omega_{\rm control} - \omega_{\rm q}$. Panels c - e show the Ramsey oscillations and fit $T_2$ coherence times at 152 h, 212 h, and 340 h from the installation of the \ce{^64}{Cu} source. During that time, the coherence time of the qubit Q1 improved from \SI{700}{\nano\second} to \SI{5.8}{\micro\second}. The average frequency of the qubit shifted by \SI{-230}{\kilo\hertz}, though this estimate is not very accurate due to the fluctuations in the qubit frequency originating from sources other than the quasiparticles, such as charge fluctuations. The observed shift can be compared to Eq. \eqref{eq:qubit_freq_xqp} with $x_{\rm qp}$ inferred from the energy-relaxation rate measurement in Fig. \ref{fig:fig2}, which yields  $\delta\omega_{\rm q}/(2\pi) = -\SI{15}{\kilo\hertz}$. The shift in the qubit frequencies is small, yet we observe a clear reduction in the coherence times of the qubit due to the ionizing radiation.
}
\bibliographystylesupp{naturemag-doi}
\bibliographysupp{references}
\end{document}